\documentclass[aps,nobalancelastpage, reprint,nofootinbib, superscriptaddress]{revtex4-2}
\usepackage{amsmath}
\usepackage{amsfonts}
\usepackage{amssymb}
\usepackage{graphicx}
\usepackage{xcolor}
\usepackage[colorlinks=True,citecolor=myBlue,linkcolor=myRed,urlcolor=myBlue]{hyperref}
\usepackage{hypcap}
\usepackage{braket}
\usepackage{yfonts}
\usepackage{bm}
\usepackage[mathscr]{eucal}
\usepackage{marginnote}

\newcommand\eq[1]{\begin{align}#1\end{align}}
\newcommand\new[1]{\textcolor{black}{#1}}

\newcommand{\M}{\overline{\mathcal{M}}}

\newcommand{\Mi}{\mathcal{M}_{i}}
\newcommand{\Ms}{\mathcal{M}_\mathrm{S}}

\newcommand{\nh}{N_\mathcal{H}}
\newcommand{\inter}{\chi_\mathrm{inter}}

\newcommand{\Fr}{\overline{F(r)}}
\newcommand{\pd}{{\phantom\dagger}}

\newcommand{\ipr}{\overline{\mathcal{L}_2}}
\newcommand{\ltwo}{\mathcal{L}_2}
\newcommand{\mbln}{\mathrm{MBL}_0}
\newcommand{\comment}[1]{}
\definecolor{myBlue}{RGB}{31,119,180}
\definecolor{myOrange}{RGB}{255,127,14}
\definecolor{myGreen}{RGB}{44,160,44}
\definecolor{myRed}{RGB}{214,39,40}
\definecolor{myPurple}{RGB}{148,103,189}

\makeatletter
\def\p@figure{\color{myBlue}}
\def\p@equation{\color{myRed}}
\makeatother

\begin{document}
\title{Fock-space \new{anatomy of eigenstates} across the many-body localisation transition}

\author{Sthitadhi Roy}
\email{sthitadhi.roy@chem.ox.ac.uk}
\affiliation{Physical and Theoretical Chemistry, Oxford University,
South Parks Road, Oxford OX1 3QZ, United Kingdom}
\affiliation{Rudolf Peierls Centre for Theoretical Physics, Clarendon Laboratory, Oxford University, Parks Road, Oxford OX1 3PU, United Kingdom}
\author{David E. Logan}
\email{david.logan@chem.ox.ac.uk}
\affiliation{Physical and Theoretical Chemistry, Oxford University,
South Parks Road, Oxford OX1 3QZ, United Kingdom}
\affiliation{Department of Physics, Indian Institute of Science, Bangalore 560 012, India}

\begin{abstract}
We explore the Fock-space \new{structure} of eigenstates across the many-body localisation (MBL) transition in a disordered, interacting quantum spin-1/2 chain.  Eigenstate expectation values of spatially local observables, which distinguish an MBL phase from an ergodic one, can be represented in terms of eigenstate amplitudes on the Fock space. Motivated by this, we introduce and study spatial correlations on the Fock space. From these, a correlation length emerges, which is found to vary discontinuously across the MBL transition; and is intimately connected to the discontinuous jump in the multifractal exponents characterising the Fock-space wavefunctions. Exploiting the direct connection between the local observables and Fock-space correlations, we show that the discontinuity in the lengthscale also implies discontinuous behaviour of the local observables across the transition. 
\new{A scaling theory based on these Fock-space correlations is constructed, which is closely connected to that for the inverse participation ratio. It yields a volume-scale in the ergodic phase and a length-scale in the MBL phase, whose critical properties suggest a Kosterlitz-Thouless-like scenario for the MBL transition, as is predicted by recent phenomenological theories.}
Finally, we also show how correlation functions on the Fock space reveal the inhomogeneities in  eigenstate amplitudes on the Fock space in the MBL phase.
\end{abstract}

\maketitle

%\tableofcontents

\section{Introduction \label{sec:intro}}

Ergodicity is an essential ingredient in the emergence of equilibrium thermodynamics from coherent quantum dynamics in generic isolated many-body systems. A key idea in this context is embodied in the eigenstate thermalisation hypothesis (ETH)~\cite{deutsch1991quantum,srednicki1994chaos,rigol2008thermalisation},  viz.\ that eigenstates of generic ergodic quantum systems locally behave like thermal states with the temperature set by their energies. In the presence of sufficiently strong quenched  disorder, however, ergodicity can be  robustly broken, leading to a many-body localised (MBL) phase~\cite{gornyi2005interacting,basko2006metal,oganesyan2007localisation,znidaric2008many} (see Refs.~\cite{nandkishore2015many,alet2018many,abanin2019colloquium} for reviews and further references). MBL systems are of fundamental importance, as they violate ETH  and hence fall outside the conventional paradigm of equilibrium 
thermodynamics and statistical mechanics. Moreover, the MBL phase at strong disorder is separated from the ergodic phase at weak disorder by an eigenstate phase transition, the precise nature of which continues to be a question of active and fundamental interest~\cite{pal2010many,luitz2015many,khemani2017critical,khemani2017two,thiery2018many-body,goremykina2019analytically,dumitrescu2018kosterlitz,roy2018exact,mace2019multifractal,morningstar2020manybody,garratt2020manybody}.

\begin{figure}[!b]
\includegraphics[width=\linewidth]{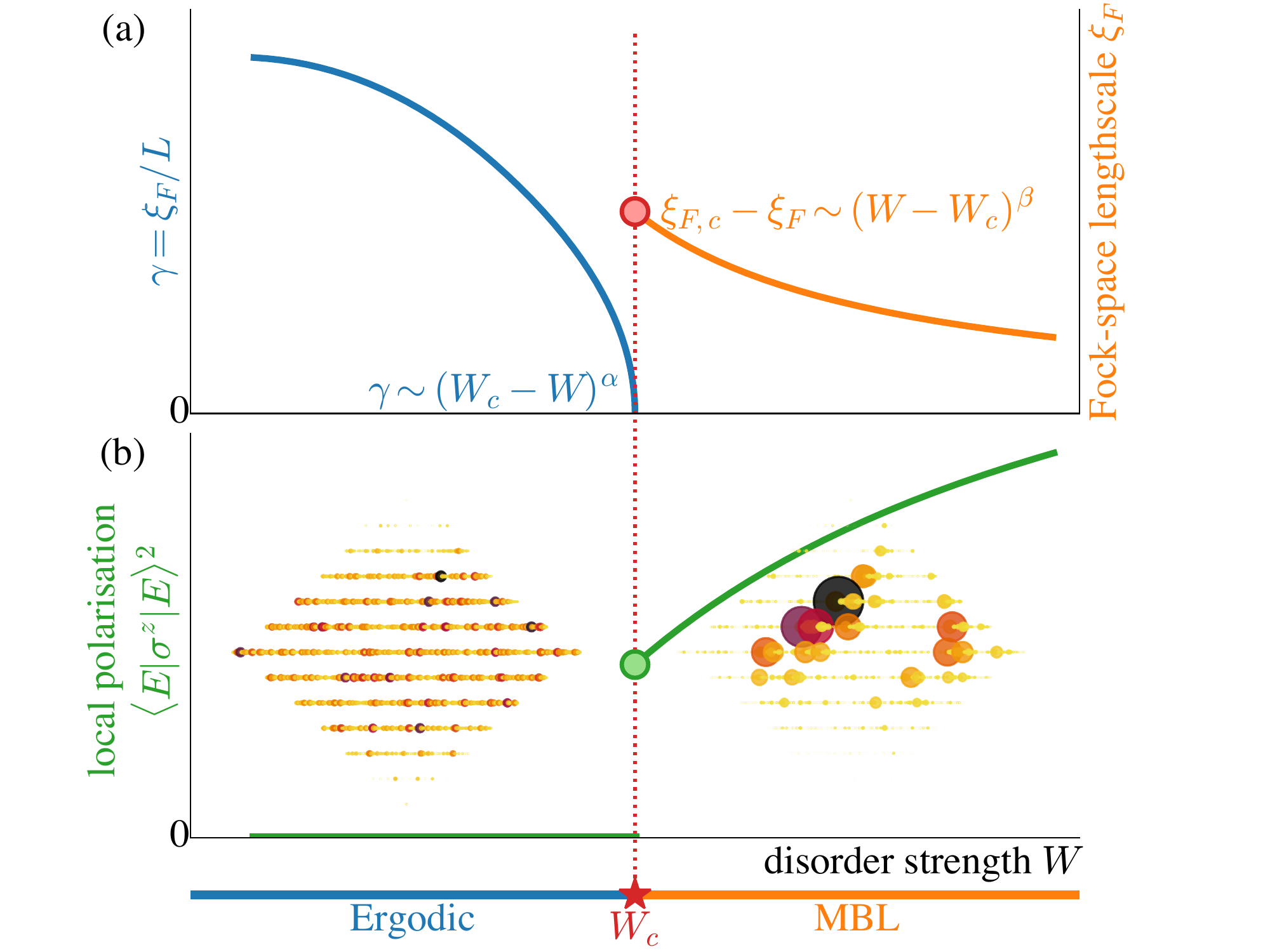}
\caption{
Schematic summary of central results. (a) There exists a correlation length $\xi_F$ on the Fock space, which is finite both throughout the MBL phase and at the MBL transition. In the ergodic phase, $\xi_F\sim \gamma L$ scales linearly with system size, with $\gamma$ vanishing at the transition $\sim (W_c-W)^\alpha$ with $\alpha\approx 0.5$. (b) This discontinuity in $\xi_F$ across the transition results in the local
polarisation, a real-space diagnostic of the transition, also being discontinuous at the transition. These results are intimately connected to the different nature of eigenstates on the Fock space in the two phases: homogeneous and extended  in the ergodic phase, inhomogeneous and fractal in the MBL phase and at the MBL transition.  This is shown graphically in the insets, where the relative size of the blobs denote the wavefunction density on the corresponding Fock-space sites. }
\label{fig:summary}
\end{figure}

The large theoretical effort towards understanding the MBL phase and accompanying transition, has seemingly forked into two complementary but 
intertwined directions. The first comprises theories, primarily phenomenological in nature, formulated directly in real space. In these approaches, the effects of ergodic spatial regions in an MBL system are treated non-perturbatively within phenomenological renormalisation group (RG) frameworks; their key predictions include the possible Kosterlitz-Thouless-like nature of the MBL transition, and an effective
real-space localisation length which is finite in the MBL phase up to and at the transition, across which it changes discontinuously~\cite{deroeck2017stability,goremykina2019analytically,dumitrescu2018kosterlitz,thiery2018many-body,morningstar2020manybody}.

The second direction, arguably more microscopically motivated, has been to study the MBL problem as an unconventional Anderson localisation problem~\cite{anderson1958absence} on the complex, correlated Fock-space graph of a quantum many-body system~\cite{altshuler1997quasiparticle,basko2006metal,MonthusGarel2010PRB,deluca2013ergodicity,serbyn2015criterion,pietracaprina2016forward,baldwin2016manybody,mace2019multifractal,logan2019many,roy2018exact,roy2018percolation,roy2019self,pietracaprina2021hilbert,detomasi2019dynamics,ghosh2019manybody,nag2019manybody,roy2020fock,biroli2020anomalous,tarzia2020manybody,detomasi2020rare,hopjan2020manybody,tikhonov2021eigenstate}. 
\new{While MBL on Fock space is inherently different from conventional Anderson localisation on high-dimensional graphs, the latter has served as an important inspiration for the former, with regard to both techniques and the scaling laws governing the transition~\cite{abou-chacra1973self,mirlin1994distribution,fyodorov1997strong,aizenman2011extended,tikhonov2016anderson,garciamata2017scaling,tikhonov2019statistics,tikhonov2019critical,garciamata2020two,tikhonov2021anderson}.}
This direction has led to crucial insights, such as the multifractal scaling of MBL eigenstates on the Fock space~\cite{deluca2013ergodicity,mace2019multifractal}, emergent fragmentation of the Fock space in the MBL phase~\cite{roy2018exact,roy2018percolation,pietracaprina2021hilbert,detomasi2019dynamics}, and the understanding that maximal correlations in the Fock-space disorder are a necessary ingredient for MBL to be stable~\cite{roy2020fock,roy2020localisation}.

Work on establishing a bridge between these two avenues, and putting them on common ground, is however fledgling. From a classical percolation viewpoint, there is an understanding of how collective effects in real space which freeze spatial segments of a system lead to a fragmentation of the Fock space~\cite{roy2018percolation}, and from a quantum mechanical viewpoint, how phenomenological distributions of such classically frozen and ergodic spatial regions can be related to the distributions of eigenstate amplitudes on the Fock space~\cite{detomasi2020rare}.

In this paper, summarised in Fig.~\ref{fig:summary}, we take a substantive step towards forging a concrete connection between the behaviour of eigenstates on the Fock space, and that of local observables across the MBL transition.  The infinite-time behaviour of dynamical autocorrelation functions, encoded in the eigenstate expectation values of local observables which diagnose the MBL transition, is shown to lead naturally to spatial correlations of eigenstate amplitudes on the Fock space, which in turn are characterised by a correlation length. The critical scaling of this correlation length is shown to be intimately connected to a scaling theory of the MBL transition  in terms of the multifractal properties of the eigenstates on the Fock space; in particular, the discontinuity in the multifractal exponent at the MBL transition~\cite{mace2019multifractal,detomasi2020rare} leads to a discontinuity in the correlation length.
\new{The correlation length is finite throughout the MBL phase as well as at the MBL critical point. On crossing the transition into the ergodic phase, the correlation length diverges discontinuously in the thermodynamic limit;
and finite-size scaling in the ergodic phase is in fact found to be controlled by a Fock-space volume-scale which diverges exponentially at the transition with an essential singularity. This is consistent with a Kosterlitz-Thouless-like scenario for the MBL transition, as is predicted by complementary approaches based on phenomenological RG in real space~\cite{deroeck2017stability,goremykina2019analytically,dumitrescu2018kosterlitz,thiery2018many-body,morningstar2020manybody}. 
The discontinuity in the Fock-space lengthscale}, via the relation between the eigenstate correlations on Fock space and the eigenstate expectation values of the local observables, is 
\new{in turn}
manifest in a discontinuity in the latter across the MBL transition.

Following an overview (Sec.~\ref{sec:overview}), the paper is organised as follows. Sec.~\ref{sec:model-loc-obs} describes the spin-1/2 model employed, and discusses results for the appropriate local observables. Eigenstate correlations in Fock space are considered in detail in Sec.~\ref{sec:fscorr}, while their connection to the Fock-space inverse participation ratios and the scaling behaviour of the latter constitutes Sec.~\ref{sec:ipr}. Section~\ref{sec:lengthscale} is dedicated to the Fock-space lengthscale associated with the eigenstate correlations. We discuss how the lengthscale emerges out of the correlation function, and its critical properties (Sec.~\ref{sec:xiF}),
together with its implications for local observables (Sec.~\ref{sec:local}), and the distributions of the correlation length 
(Sec.~\ref{sec:xiFdist}).  Sec.~\ref{sec:inhomo} presents results for the finer-grained inhomogeneous structure of the eigenstates across Fock space in the MBL phase, and its absence in the ergodic phase.  Concluding remarks are given in Sec.~\ref{sec:conclusion}.

\tableofcontents

%%%%%%%%%%%%%%%%%%%%%%%%%%%%%%%%%%%%%%%%%%%%%%%%%%%%%%%%%%%%%%%%%%%%%%%%%%%%%%%%%%%%%%%%%%%%

\subsection{Overview \label{sec:overview}}
The central results of this work can be stated succinctly in the following two points
(see also Fig.~\ref{fig:summary}):
\begin{itemize}
	\item Eigenstate expectation values of local observables which diagnose the MBL phase and transition, can be expressed in terms of spatial correlations between eigenstate amplitudes on the Fock-space graph. 
	\item The associated correlation length changes discontinuously across the MBL transition. It is finite throughout the MBL phase and at the critical point, but is divergent throughout the ergodic phase. This discontinuity manifests itself in a discontinuous behaviour of the local observables across the transition.
\end{itemize}

We consider specifically a disordered quantum spin-1/2 chain, where the disordered fields and interactions couple to the $\sigma^z$-component of the spins. The relevant local observables to study are thus the real-space dynamical autocorrelations and eigenstate expectation values of the local $\sigma^z$-operators. At the same time, from a Fock-space perspective, it is natural to consider the basis of $\sigma^z$-product states, to which the MBL eigenstates are smoothly connected in the  strong-disorder limit. Within this setting, we show that the average infinite-time value of the $\sigma^z$-autocorrelation measured in an eigenstate, is directly related to a spatial two-point correlation function defined on the Fock space, between eigenstate amplitudes on Fock-space sites. For an eigenstate $\ket{E}$ decomposed in terms of the Fock-space basis states $\{\ket{I}\}$ as $\ket{E}=\sum_{I}A_I\ket{E}$,  the correlation function $F(r)$ is defined as 
\eq{
F(r)=\sum_{I,K:r_{IK}^{\pd}=r}|A_I|^2|A_K|^2\,,\nonumber
}
with $r_{IK}$ the Hamming distance between sites $I$ and $K$. The infinite-time autocorrelation is related to the Fock-space correlation function by
\eq{
\lim_{t\to\infty}\frac{1}{L}\sum_{i=1}^L\braket{E|\sigma^z_i(t)\sigma^z_i|E}=\sum_{r=0}^L\left(1-\frac{2r}{L}\right)F(r)\,,\nonumber
}
with $L$ the system size. This relation provides a bridge between the behaviour of local observables in real space and spatial correlations of eigenstates on the Fock space.

We show, both numerically and constructively, that  $F(r)$ is characterised by a correlation length, $\xi_F$, on the Fock space. Since the MBL transition is a bona fide phase transition on the Fock-space graph, the critical scaling of $\xi_F$ is  a question of  central interest. To 
address this, we exploit the fact that the correlation function $F(r=0)$ (which itself encodes $\xi_F$) is the Fock-space inverse participation ratio (IPR) of the eigenstate. To that end we first study a scaling theory for the transition in terms of the Fock-space IPRs (or participation entropies (PE)), in direct parallel to that for a disordered XXZ chain~\cite{mace2019multifractal}; and then use it to extract the critical  properties of $\xi_F$.

The scaling theory in terms of IPRs or PEs (Ref.~\cite{mace2019multifractal} and Sec.~\ref{sec:ipr})  tells us that the entire MBL phase as well as the MBL critical point is characterised by non-ergodic multifractal eigenstates.  The IPRs follow a scaling function of $\ln \nh/\xi$,
with $\nh$ the Fock-space dimension, and $\xi$  an emergent lengthscale which diverges as $\xi\sim (W-W_c)^{-\beta}$ with $W$ the disorder strength. The implication of this for $\xi_F$ is shown to be that it is finite  throughout the MBL phase, and approaches a finite limit $\xi_{F,c}$ at the transition as $\xi_{F,c}-\xi_F\sim (W-W_c)^\beta$, with the same exponent  $\beta$ [see Fig.~\ref{fig:summary}(a)].   Although $\xi_F$ is finite in the MBL phase, the multifractality of the eigenstates arises from competition between the exponential decay of eigenstate correlations between Fock-space sites at mutual Hamming distance $r$  (with $\xi_F$ the decay lengthscale), and the exponential growth of the number of Fock-space sites at distance $r$ from any given site.

On the ergodic side of the transition by contrast, the scaling theory shows that in the thermodynamic limit the eigenstates are fully ergodic throughout the phase, and  that the multifractal exponent jumps discontinuously at the transition.  Here, the IPRs follow a scaling function of $\nh/\Lambda_{2}$, where $\Lambda_{2}$ can be understood as a non-ergodic volume~\cite{mace2019multifractal}, and diverges as $\Lambda_{2}\sim \exp[(W_c-W)^{-\alpha}]$ with $\alpha\approx 0.5$. This is shown to result in $\xi_F$ scaling with the system size $L$ as $\xi_F=\gamma L$, where $\gamma$ vanishes at the transition with the same exponent, $\gamma\sim(W_c-W)^\alpha$ [see Fig.~\ref{fig:summary}(a)].
The discontinuous behaviour in $\xi_F$ across the transition in turn leads to a discontinuity in the infinite-time autocorrelation function of the local $\sigma^z$,
\eq{
 \lim_{t\to\infty}\frac{1}{L}\sum_{i=1}^L\braket{E|\sigma^z_i(t)\sigma^z_i|E}=1-2\left(1+e^{1/\xi_F}\right)^{-1}\,.
	\nonumber
}
This vanishes in the thermodynamic limit throughout the ergodic phase, jumps discontinuously to a finite value at the critical point, and
thereafter in the MBL phase grows continuously towards $1$ with increasing $W$ [see Fig.~\ref{fig:summary}(b)].

An essential physical intuition behind the IPR-based scaling theory is that the MBL eigenstates reside on sparse `strands' on the Fock-space graph. We further show how generalisations of the correlation function $F(r)$ probe these finer structures of the eigenstates, and reveal that they are indeed strongly inhomogeneous. The inhomogeneities probed by these generalised correlation functions go well beyond those probed by the multifractal scaling of the IPRs.

%%%%%%%%%%%%%%%%%%%%%%%%%%%%%%%%%%%%%%%%%%%%%%%%%%%%%%%%%%%%%%%%%%%%%%%%%%%%%%

\section{Model and local observables \label{sec:model-loc-obs}}
\subsection{Disordered spin-1/2 chain}
To place our discussions on a concrete footing, we employ a disordered spin-1/2 chain which hosts a firmly established MBL phase~\cite{imbrie2016many}. It is specified by the Hamiltonian
\eq{
H = \sum_{i=1}^{L-1}J_i^{\pd}\sigma^z_{i}\sigma^z_{i+1}+ \sum_{i=1}^L[h_i^{\pd}\sigma^z_i + \Gamma\sigma^x_i]\,,
\label{eq:ham}
}
where $J_i$ and $h_i$ are uniformly distributed random numbers with $J_i\in[J-\delta J,J+\delta J]$ and $h_i\in[-W,W]$.  In the numerical studies employed we focus solely on eigenstates in the middle of the spectrum, and for each disorder realisation use a single eigenstate, $\ket{E}$, with its energy closest to $\mathrm{Tr}[H]=0$. We consider throughout $J=1$, $\delta J=0.2$, and $\Gamma=1$. For these parameters, the MBL transition for the model occurs at a critical disorder strength $W_c\simeq 3.5$~\cite{abanin2021distinguishing}, determined from level statistics and bipartite entanglement entropy.

%%%%%%%%%%%%%%%%%%%%%%%%%%%%%%%%%%%%%%%%%%%%%%%%%%%%%

\subsection{Eigenstate polarisation and autocorrelation}

 A defining signature of the MBL phase is a persistent local memory of the initial conditions throughout the course of time-evolution. This is often quantified via local temporal autocorrelations measured with respect to the eigenstates, and their infinite-time values. Since the disorder couples to $\sigma^z$-components of the spins-1/2 in the model Eq.\  \ref{eq:ham}, the relevant autocorrelation is
\eq{
	\mathcal{A}_i^\pd(t) = \braket{E|\sigma^z_i(t)\sigma^z_i|E}\,,
}
the infinite-time limit of which is 
\eq{
	\mathcal{M}_i^{\pd} \equiv \lim_{t\to\infty}\mathcal{A}_i^\pd(t) = \braket{E|\sigma^z_i|E}^2\,.
	\label{eq:Mi}
}
For a given disorder realisation, we denote the average of $\mathcal{M}_i$ over all sites as 
$\mathcal{M}_\mathrm{S}$, and the latter's disorder average by $\overline{\mathcal{M}}$,
\eq{
\mathcal{M}_\mathrm{S}^{\pd} = L^{-1}\sum_{i=1}^L\mathcal{M}_i, ~~~~~~
\overline{\mathcal{M}}=\overline{\mathcal{M}_{S}}\,. 
\label{eq:Miplus}
}

In an MBL phase the local autocorrelation saturates to a system-size independent finite value at infinite times, whereas in an ergodic phase it decays to zero in the thermodynamic limit. Equivalently, via Eq.~\ref{eq:Mi}, $\overline{\mathcal{M}}$ encodes how strongly the spins are polarised along the $z$-direction in the eigenstates.

\begin{figure}
\includegraphics[width=\linewidth]{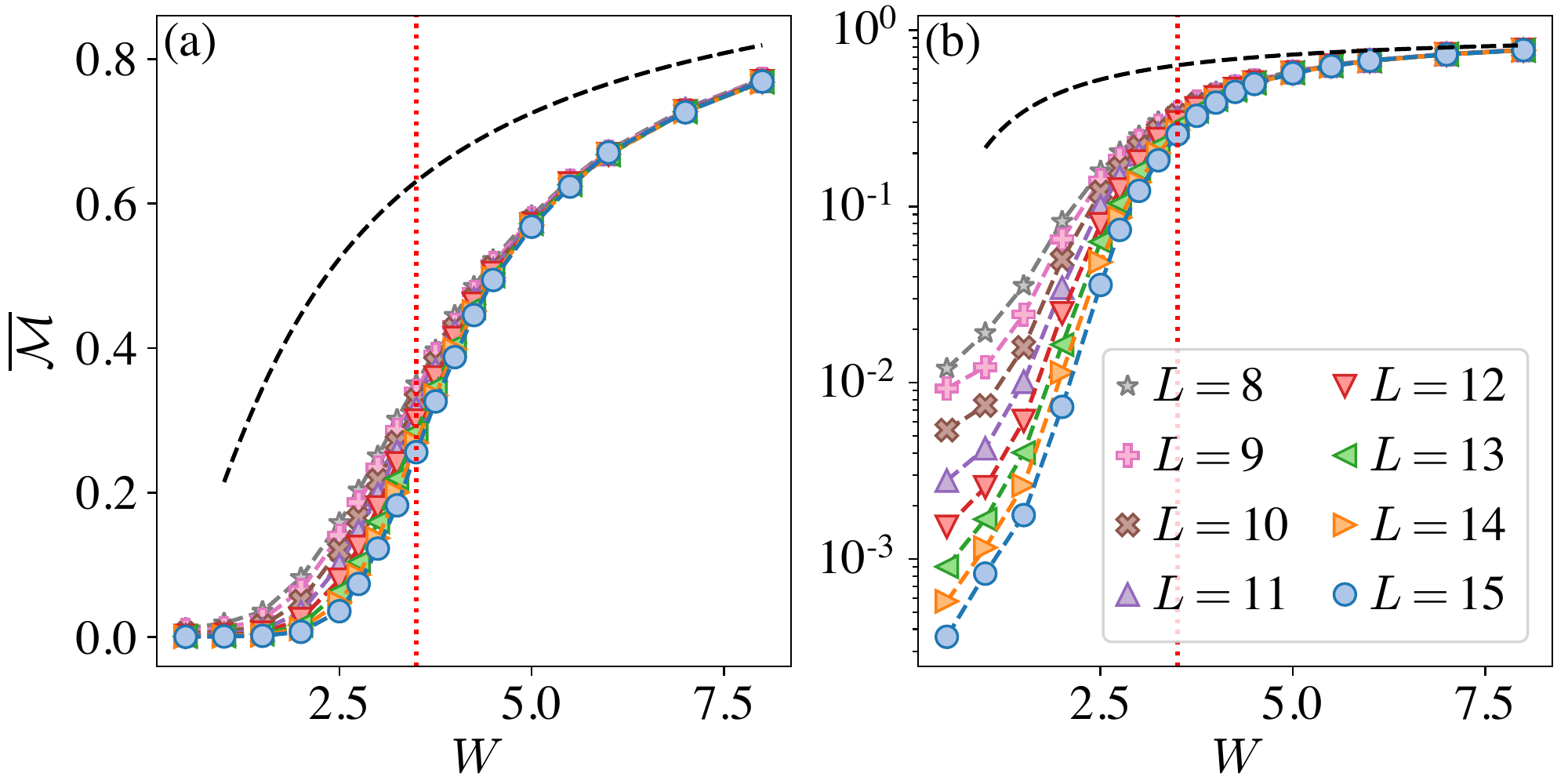}
\caption{Disorder-averaged eigenstate polarisation $\overline{\mathcal{M}}$ (Eq.~\ref{eq:Miplus}) \emph{vs} disorder strength $W$ for different system sizes $L$. Panels (a) and (b) show data on linear and logarithmic scales respectively. The vertical red dotted line is a guide to the eye for the MBL transition at $W_c\simeq3.5$~\cite{abanin2021distinguishing}. The black dashed line denotes the result in the $\mbln$ case of $J_i=0$ (Appendix~\ref{sec:non-int-corr}).
}
\label{fig:M}
\end{figure}

Numerical results for $\overline{\mathcal{M}}$ are shown in Fig.~\ref{fig:M}. Deep in the ergodic phase, the finite-size scaling of the ETH~\cite{beugeling2014finitesize} suggests that $\braket{E|\sigma^z_i|E}$ is normally distributed with a standard deviation $\propto \nh^{-1/2}$, implying that $\M\propto \nh^{-1}$. This is indeed reflected in the data in panel (b). On a logarithmic scale data for different $L$ are equispaced, indicating that $\M$ decays as a power of $\nh$, and hence exponentially with $L$. The MBL phase by contrast is characterised by a finite $L$-independent $\overline{\mathcal{M}}$, which is also reflected well in the data shown in Fig.~\ref{fig:M}.  The behaviour of $\M$ with system size is likewise consistent with $W_c\simeq 3.5$~\cite{abanin2021distinguishing}. For $W\lesssim3.5$, $\M$ decays systematically with $L$ towards zero, while for $W\gtrsim3.5$ it appears to saturate to a finite value independent of $L$ as expected in an MBL phase. Note that the data also hints rather strongly at a discontinuous jump of $\M$ at the MBL transition; we will return to this issue at length later.

Further insight into the behaviour of the local polarisations Eq.\ \ref{eq:Mi}  can be obtained from their probability distributions. In particular, we consider two distributions. The first, defined as,
\eq{
	P_{\Mi}^{\pd}(m) = \overline{L^{-1}{\textstyle\sum_{i=1}^{L}}\delta(\Mi-m)}\,
	\label{eq:Pmi}
}
is a distribution of the local polarisation over both real-space sites and disorder realisations. The second, defined by,
\eq{
	P_{\Ms}^{\pd}(m) = \overline{\delta(\Ms-m)}=\overline{\delta(L^{-1}{\textstyle\sum_{i=1}^{L}}\Mi-m)}\,
	\label{eq:Pms}
}
is the distribution of the sample-averaged polarisation, $\Ms$, over disorder realisations.

\begin{figure}
\includegraphics[width=\linewidth]{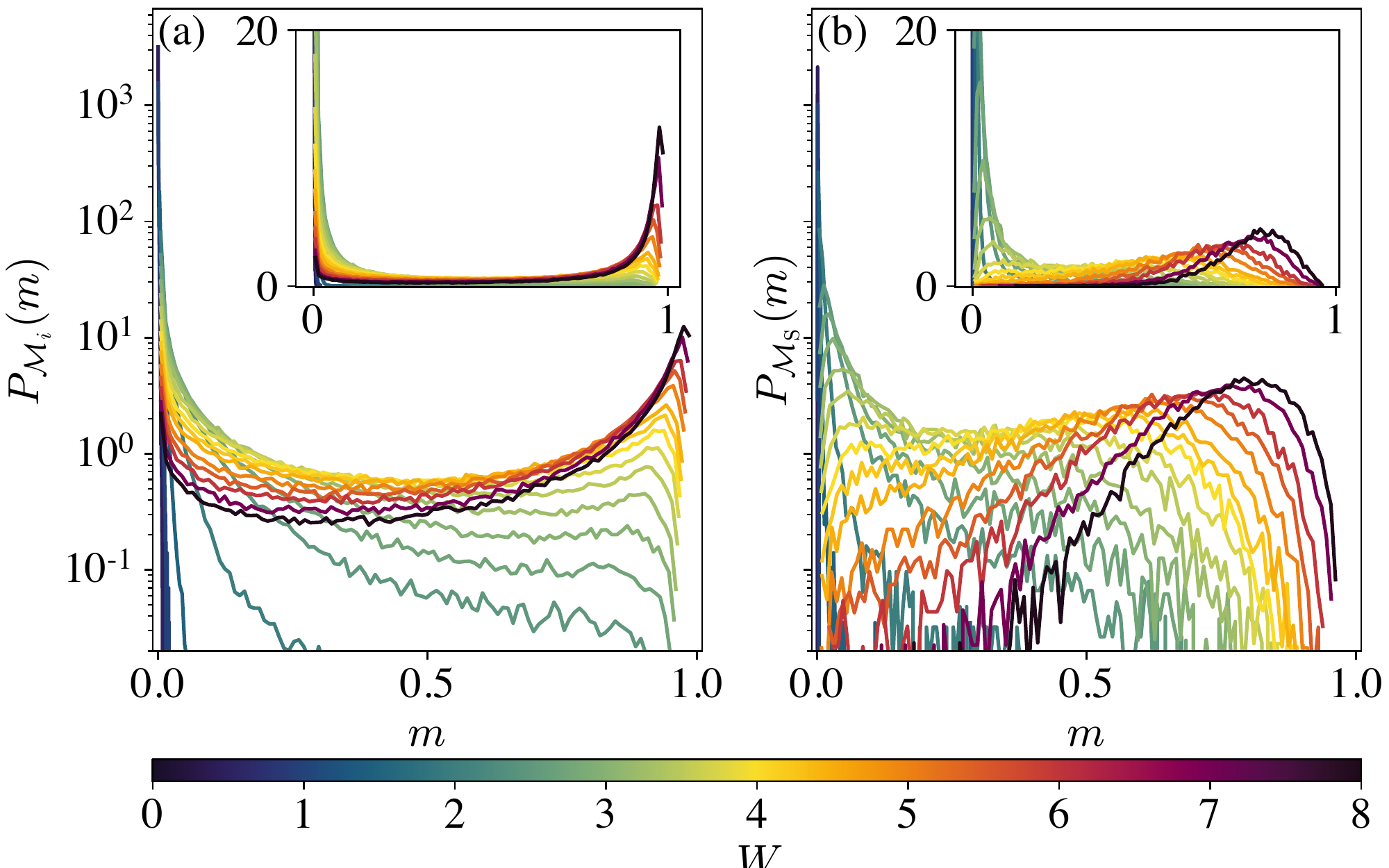}
\caption{Probability distributions of the eigenstate polarisation for different disorder strengths $W$. Panel (a) shows distributions of the local polarisation, $\mathcal{M}_i$, over both real-space sites and disorder realisations (Eq.~\ref{eq:Pmi}). Panel (b) shows the distribution 
Eq.~\ref{eq:Pms} over disorder realisations of the site-averaged polarisation $\mathcal{M}_\mathrm{S}$. Insets show same distributions as main panel, but on a linear scale. Data shown is for $L=15$. }
\label{fig:dists}
\end{figure}

Numerical results for the two distributions are shown in Fig.~\ref{fig:dists}. In the ergodic phase, both $P_{\Mi}$ and $P_{\Ms}$ are sharply peaked near zero and the distributions are rather narrow, consistent with ETH~\cite{beugeling2014finitesize}. At intermediate disorder and near the MBL transition, both distributions become broad. Note that we refer here to the distribution as being broad if it has a finite second central moment in the thermodynamic limit, and narrow otherwise (the support of the distributions is strictly compact $\in[0,1]$, so none of the moments can diverge).

A qualitative difference between the two distributions emerges in the MBL phase. While $P_{\Mi}$ remains broad with a peak appearing at $m\lesssim 1$, $P_{\Ms}$ becomes narrow again,  as indicated by the exponential decay of the distribution away from its own peak at $m\lesssim 1$.
The broadness in $P_{\Mi}$ throughout the MBL phase can be understood physically as follows. In any disorder realisation, there would be a finite density of spins which locally experience a disorder strength weaker than the critical one. These spins thus attempt to thermalise, and contribute to the weight of the distribution $P_{\Mi}$ at $m\sim 0$. The  remaining spins on the other hand retain their $\sigma^z$-polarisations to various degrees, and as such populate the rest of the support of $P_{\Mi}(m)$ for $m\in (0,1]$. The distribution $P_{\Mi}$ is therefore broad. Note that the same picture was borne  out by the Fock-space percolation proxy of the MBL transition~\cite{roy2018exact,roy2018percolation}, and an argument based on the picture was used to rationalise the fractal nature of the eigenstates on Fock space in the MBL phase~\cite{mace2019multifractal,detomasi2020rare}. On the other hand, the distribution $P_{\Ms}$ of the site-averaged polarisation appears to be
narrow deep inside either phase, but broad in the critical regime. This implies that $\Ms$ is self-averaging in either of the two phases, but not so near the critical point.

The above features of the distributions can be quantified via their second central moments. The fluctuation in $\Mi$ over both real-space sites and disorder realisations is defined as 
\eq{
	\chi = L^{-1}{\textstyle\sum_{i=1}^L} \overline{\Mi^2} - \left(L^{-1}\textstyle{\sum_{i=1}^L} \overline{\Mi}\right)^2\,,
	\label{eq:chi}
}
which is simply the second central moment of $P_{\Mi}$; while the second central moment of $P_{\Ms}$ is
\eq{
	\inter = \overline{\Ms^2}-(\M)^2\,,\label{eq:chi-inter}
}
and physically quantifies the fluctuation over disorder realisations of the sample-averaged local polarisation $\Ms$. 

\begin{figure}
\includegraphics[width=\linewidth]{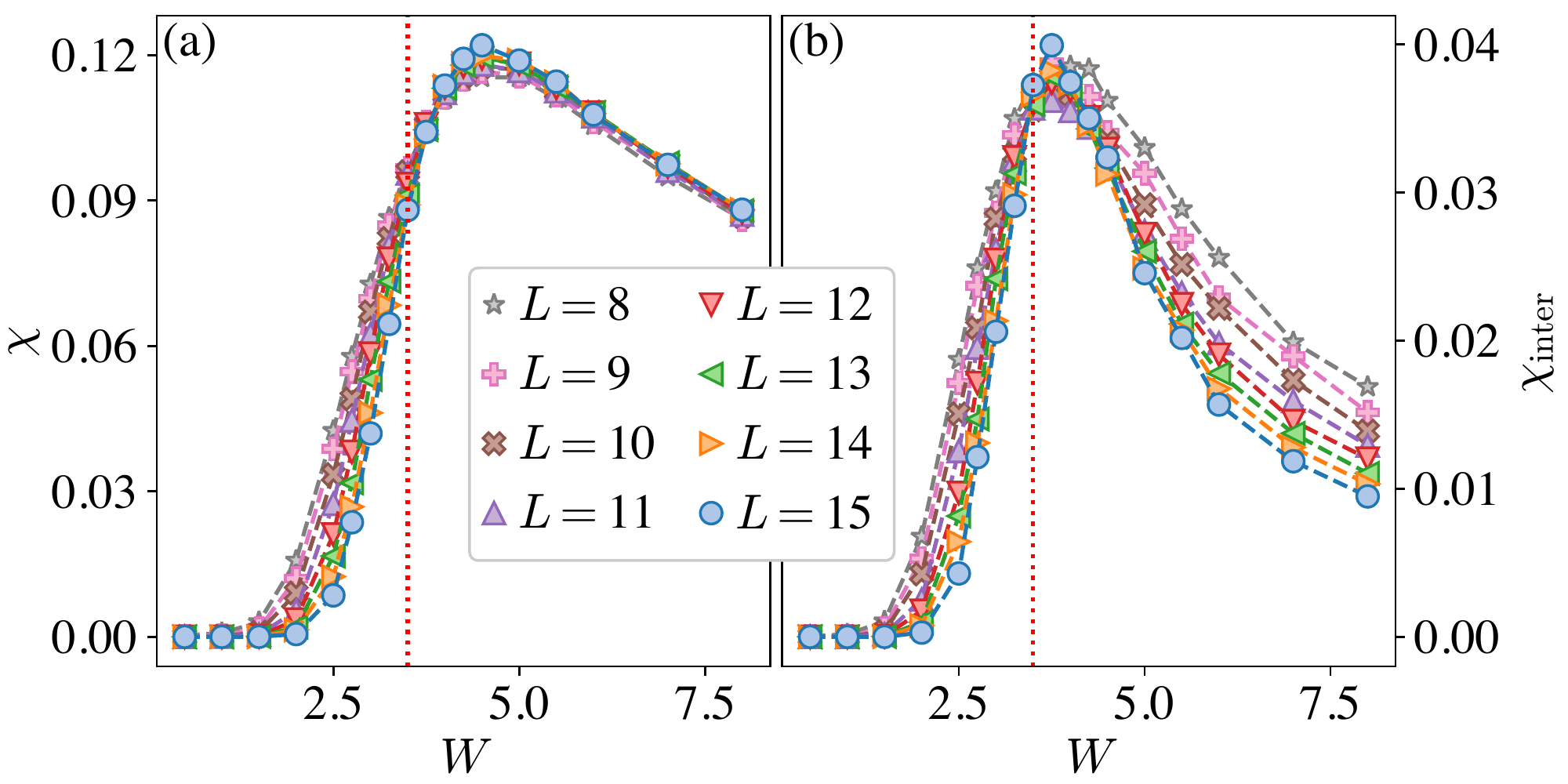}
\caption{The total and inter-sample fluctuations in the eigenstate polarisation, $\chi$ [panel (a)] and $\inter$ [panel (b)], defined via Eqs.~\ref{eq:chi},\ref{eq:chi-inter}, \emph{vs} disorder strength $W$ for different system sizes $L$. $\chi$ decays systematically with $L$ in the ergodic phase and tends to a finite value with increasing $L$ in the MBL phase. $\inter$ decays systematically with $L$ in both phases, but shows a finite $L$-independent peak at the transition.}
\label{fig:chi}
\end{figure}

Fig.~\ref{fig:chi} shows numerical results for $\chi$ and $\inter$. In the ergodic phase both $\inter$ and $\chi$ decay to zero with $L$ as mandated by ETH. This  reflects the fact that both distributions, $P_{\Mi}$ and $P_{\Ms}$, are narrow in the ergodic phase. In the MBL phase by contrast, while $\inter$ vanishes with increasing $L$, $\chi$ saturates to a finite value. This is consistent with $P_{\Ms}$ and $P_{\Mi}$ being respectively narrow and broad in the MBL phase (see Fig.~\ref{fig:dists}). Although $\inter$ vanishes with increasing $L$  within each phase, it appears to saturate to an $L$-independent peak at the critical point (the residual $L$-dependence seen in Fig.~\ref{fig:chi}(b) is not systematic with $L$). This reflects the broadness of $P_{\Ms}$ and the absence of self-averaging in $\Ms$ at the critical point. As evident in Fig.~\ref{fig:chi}(b), this $L$-independent peak is in excellent agreement with the estimated critical $W_c$  (vertical dotted 
line). We  also add that the peak in $\chi$ itself (Fig.~\ref{fig:chi}(a)) is of no significance, as $\chi$ is expected to be respectively finite and vanishing throughout the MBL and ergodic phases. Indeed, the data in Fig.~\ref{fig:chi}(a) shows that the $L$-dependence of $\chi$ is consistent with the estimated $W_c\simeq 3.5$~\cite{abanin2021distinguishing}; for $W<W_c$, $\chi$ decays systematically with $L$, whereas for $W>W_c$ it is $L$-independent. 

%%%%%%%%%%%%%%%%%%%%%%%%%%%%%%%%%%%%%%%%%%%%%%%%%%%%%%%%%%%%%%%%%%%%%%%%%%%%%%%%%%%%%%%%%%%%%%%%%%%%%%%%%%%

\section{Spatial correlations of eigenstates on Fock space \label{sec:fscorr}}
Having established the behaviour of the local polarisations across the MBL transition, we now turn to spatial correlations in the eigenstates on the Fock space, and show how they encode the local polarisations and signatures of the MBL transition in general.

\begin{figure}
\includegraphics[width=\columnwidth]{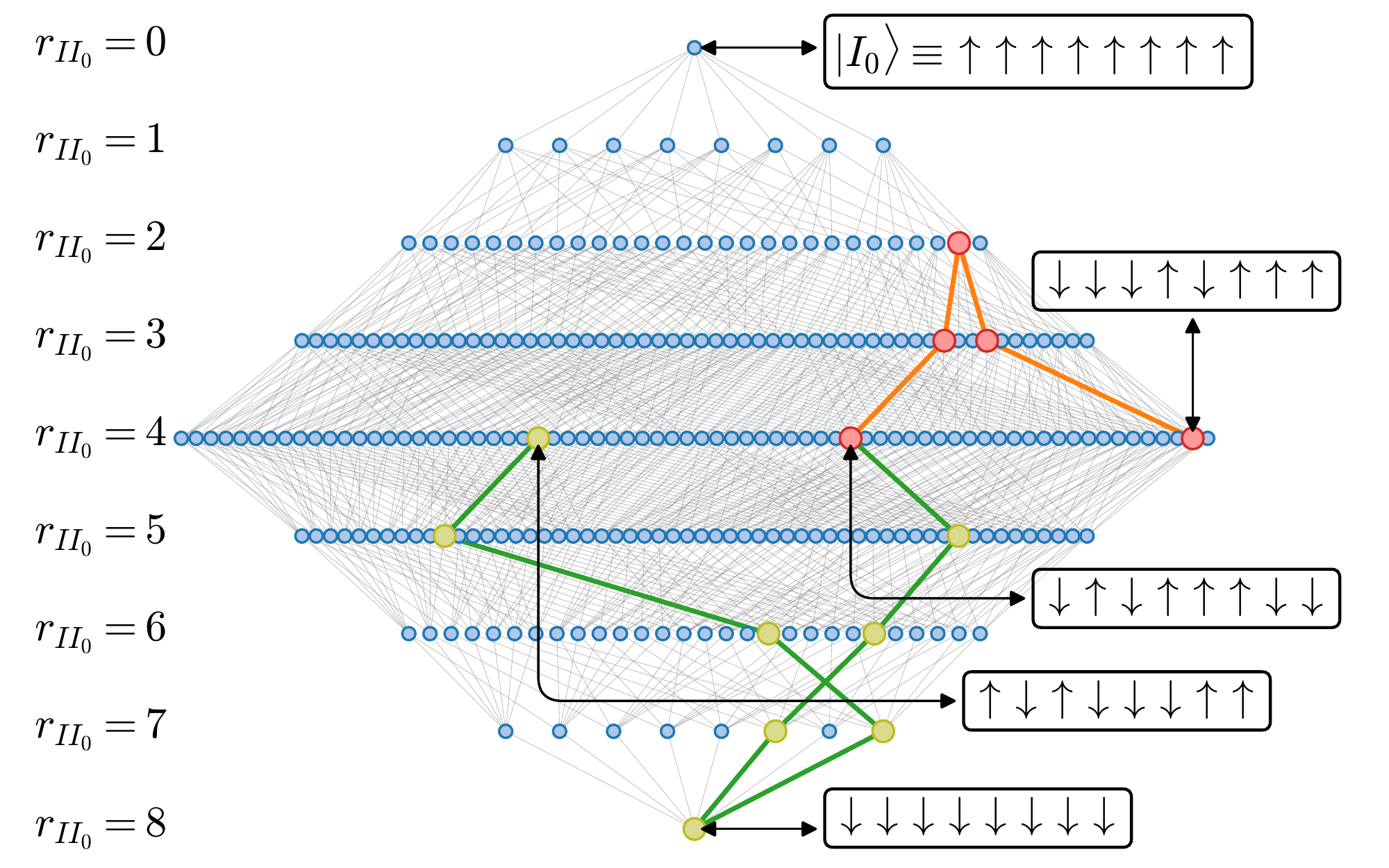}
\caption{Fock-space graph of the disordered Ising chain in Eq.~\ref{eq:ham} in the basis of $\sigma^z$-product states, with $L=8$.
A Fock-space site, denoted by $\ket{I_0}$ and here chosen to have all spins up, is placed at the apex; and any Fock-space site $I$ with $r_{II_0}=r$ is a Hamming distance $r$ from $I_0$. Note that two sites with the same $r_{II_0}$ can lie at a distance $2s$ from each other, where $s=0,1,\cdots,\min(r,L-r)$. Representative examples are shown by the red and yellow coloured sites.}
\label{fig:fockspace}
\end{figure}

The Fock-space of the model Eq.~\ref{eq:ham} in the basis of $\sigma^z$-product states is an $L$-dimensional hypercube with $\nh=2^L$ vertices, as illustrated in Fig.~\ref{fig:fockspace}. A vertex, denoted as $\ket{I}$, is a many-body quantum state of $L$ spins, which is an eigenstate of the $\sigma^z_i$-operator, $\sigma^z_i\ket{I}=S_{i,I}\ket{I}$, where $S_{i,I}=\pm 1$. Since the links on the Fock-space graph are generated by the term $\sum_i\sigma^x_i$ in the Hamiltonian, each Fock-space site is connected to  precisely $L$ others, each corresponding to flipping the spin at a particular real-space site.

An eigenstate, $\ket{E}$, can be decomposed on the Fock-space as  $\ket{E}=\sum_I A_I\ket{I},$ with $\sum_I |A_I|^2=1$  from normalisation (while the amplitudes $A_I$  depend on the eigenstate $\ket{E}$, we omit it notationally since we always focus on a single eigenstate). As a measure of distance between two vertices on the Fock-space graph we use the Hamming distance, which is the number of real-space sites on which the spin orientation differs between the  vertices. For the specific model Eq.~\ref{eq:ham} considered, the Hamming distance is equivalently the shortest distance between the vertices. The distance between two vertices (Fock-space sites henceforth), $\ket{I}$ and $\ket{K}$, can be expressed as
\eq{
	r_{IK}^\pd = \frac{1}{4}\sum_{i=1}^L (S_{i,I}-S_{i,K})^2\,,
	\label{eq:rIK}
}
or equivalently $L^{-1}\sum_{i}S_{i,I}S_{i,K}=1-2r_{IK}/L$. This relation between the distance on the  Fock-space graph and the spin orientations plays a central role in connecting the Fock-space landscape of eigenstates to the local polarisations measured with respect to them.

%%%%%%%%%%%%%%%%%%%%%%%%%%%%%%%%%%%%%%%%%%%%%%%%%%%%%%%%%%%%%%%%%%%%%%%%%%%%%%%%%%%%%

\subsection{Eigenstate correlations and local polarisations}
\begin{figure}[!t]
\includegraphics[width=\linewidth]{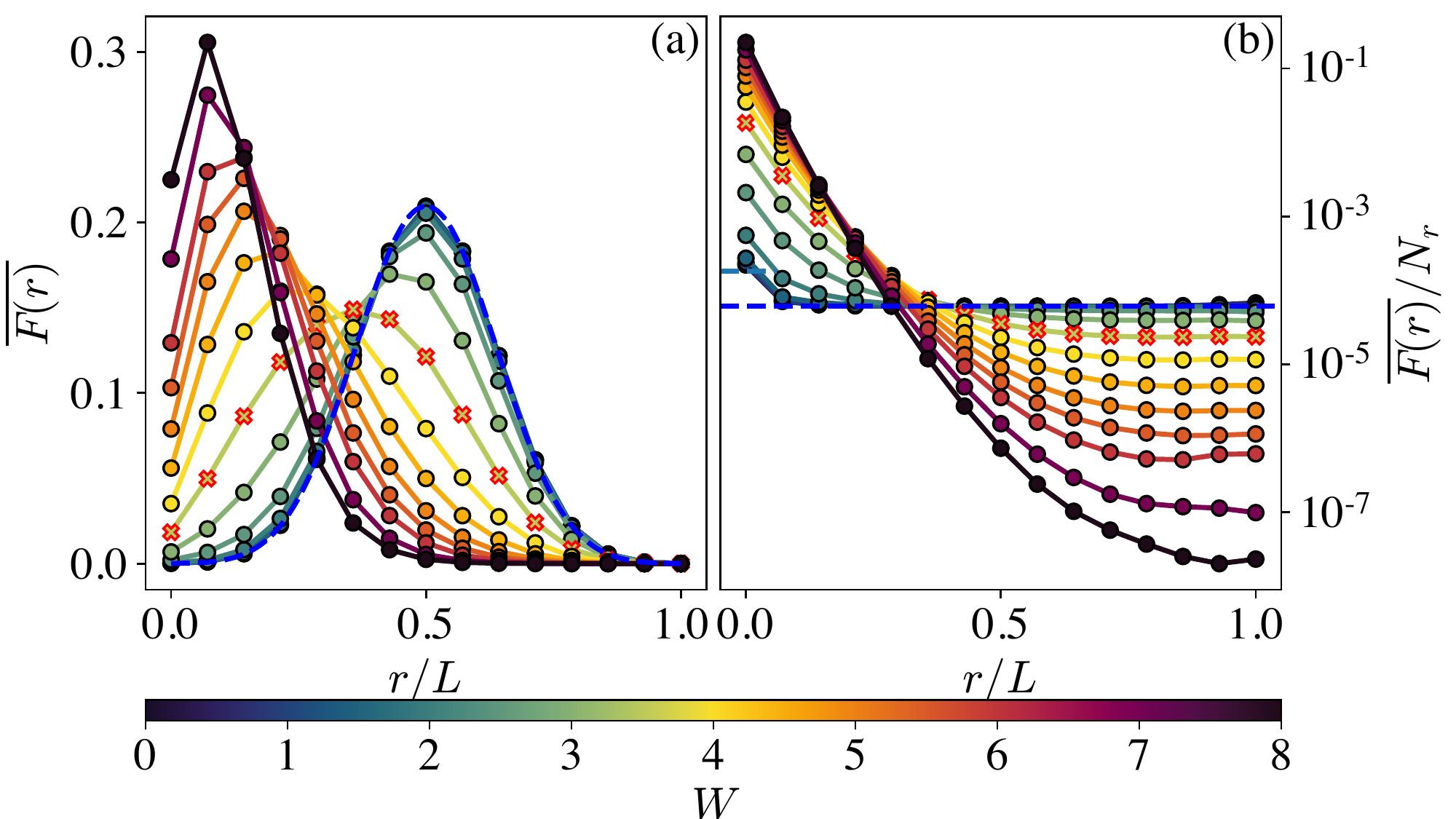}
\caption{(a) Spatial correlation of eigenstates on the Fock space, $\overline{F(r)}$ (see Eq.~\ref{eq:Fr}), \emph{vs} $r/L$ for various disorder strengths $W$.  Dark blue dashed line denotes the binomial distribution $B(L,p=1/2)$. (b) Same data as in (a) but rescaled with $N_r=\binom{L}{r}$.  Light and dark blue horizontal dashed lines correspond respectively to $3/\nh$ and $1/\nh$, in agreement with  Eq.~\ref{eq:Ferg}. Data marked with red crosses correspond to the critical   $W_c=3.5$. Results are shown for $L=14$.}
\label{fig:Fr}
\end{figure}

The local polarisation $\mathcal{M}_{S}$, a real-space diagnostic of the MBL transition, can be expressed in terms of the Fock-space amplitudes as
\eq{
	\Ms &= L^{-1}\sum_{i=1}^L \sum_{I,K}\vert A_I^{\pd}\vert^2\vert A_K^{\pd}\vert^2 S_{i,I}^{\pd}S_{i,K}^{\pd}\nonumber\\
	&=\sum_{I,K}\vert A_I^{\pd}\vert^2\vert A_K^{\pd}\vert^2 (1-2r_{IK}^{\pd}/L)\,
	\label{eq:Ms-r1}
}
(using Eq.~\ref{eq:rIK} in the second line). Eq.~\ref{eq:Ms-r1} motivates a spatial correlation on the Fock space for an eigenstate as
\eq{
	F(r) = \sum_{I,K: r_{IK}^{\pd}=r} |A_I|^2|A_K|^2\,,
	\label{eq:Fr}
}
in terms of which the local polarisation $\Ms=\sum_r F(r)(1-2r/L)$. The correlation function $F(r)$  is closely connected to the non-local propagator on the Fock space,  $G_{IK}(\omega) = \braket{I|(\omega+i0^{+}-H)^{-1}|K}$, by
\eq{
	F(r) = \sum_{I,K: r_{IK}^{\pd}=r} |\mathrm{Res}_E^{\pd} G_{IK}^{\pd}(\omega)|^2\,,
	\label{eq:Fr-GIK}
}
with $\mathrm{Res}_E G_{IK}(\omega)$ the residue of the $G_{IK}(\omega)$ at  eigenenergy $E$. The normalisation of the eigenfunctions leads to $\sum_{r=0}^L F(r)=1$, so that $F(r)$ can be interpreted as a normalised probability distribution over $r$. This allows us to define a mean distance on the Fock space as 
\eq{
	\braket{r} = \sum_{r=0}^L r F(r)\,.
	\label{eq:rmean}
}
At the same time, the real-space site-averaged polarisation $\Ms$ can expressed in terms of $\braket{r}$ using Eqs.~\ref{eq:Ms-r1}, \ref{eq:Fr}, and \ref{eq:rmean} as
\eq{
	\Ms =1-2\frac{\braket{r}}{L}\,.
	\label{eq:Ms-r}
}
This provides a direct relation between the mean local polarisation in real space, and the structure of eigenstates on Fock space. The variance of the distribution $F(r)$ can also be defined as 
\eq{
	\braket{(\delta r)^2} = \sum_{r=0}^L r^2F(r) - \braket{r}^2\,.
	\label{eq:rsq-mean}
}

\begin{figure}[!t]
\includegraphics[width=\linewidth]{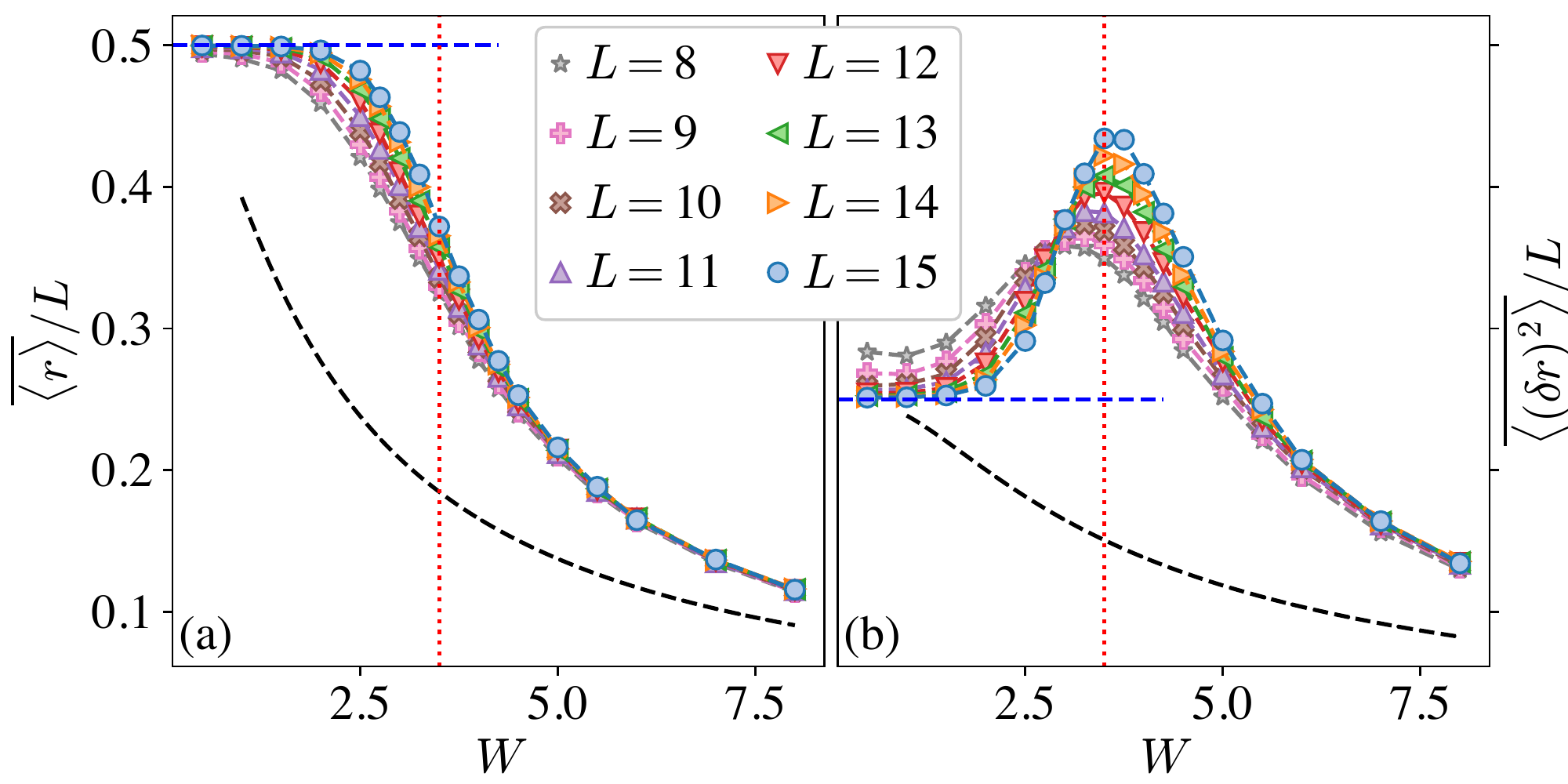}
\caption{Mean distance $\overline{\braket{r}}$ and its variance $\overline{\braket{(\delta r)^2}}$,  Eqs.~\ref{eq:rmean} and \ref{eq:rsq-mean} respectively,  \emph{vs} disorder strength $W$ for various systems sizes $L$.  Blue horizontal dashed lines correspond to the ergodic values of $\overline{\braket{r}}/L=1/2$ and $\overline{\braket{(\delta r)^2}}/L=1/4$.  Black dashed lines show results for the $\mbln$ limit of $J_i=0$.}
\label{fig:r-rsq}
\end{figure}

Numerical results for $F(r)$ are shown in Fig.~\ref{fig:Fr}. Deep in the ergodic phase, the eigenstate amplitudes $A_I$ are independent and normally distributed, with zero mean and a variance $\nh^{-1}$. Using this in Eq.~\ref{eq:Fr} gives
\eq{
	\overline{F_\mathrm{erg}(r)} = \frac{3}{\nh}\delta_{r,0}^{\pd}+(1-\delta_{r,0}^{\pd})\frac{N_r}{\nh}\,,
	\label{eq:Ferg}
}
with $N_r=\binom{L}{r}$ the number of Fock-space sites at distance $r$ from any site on the Fock-space graph. As $\nh=2^L$, $\overline{F_\mathrm{erg}(r)}$ in Eq.~\ref{eq:Ferg} is equivalent to a binomial distribution $B(L,p)$ with $p=1/2$, except for the $\delta_{r,0}\nh^{-1}$ correction.\footnote{A random number $x\sim B(L,p)$ distributed according to a binomial distribution has a probability distribution function $\binom{L}{x}p^x(1-p)^{L-x}$.} As indicated by the blue dashed line in Fig.~\ref{fig:Fr}(a), this binomial distribution is indeed in excellent agreement with the data at weak disorder. Concomitantly, when scaled by $N_r$  (as in Fig.~\ref{fig:Fr}(b)), the data agree perfectly with the prediction of Eq.~\ref{eq:Ferg}, including the correction at $r=0$. Using this form of $F(r)$ in Eqs.~\ref{eq:rmean} and \ref{eq:rsq-mean}, gives
\eq{
	\frac{\overline{\braket{r}}_\mathrm{erg}}{L}=\frac{1}{2};\quad\frac{\overline{\braket{(\delta r)^2}}_\mathrm{erg}}{L} = \frac{1}{4}\,
	\label{eq:r-erg}
}
in the ergodic phase. This is also evident in the numerical results presented in Fig.~\ref{fig:r-rsq}. With increasing $L$ the data tend towards the values in Eq.~\ref{eq:r-erg} not only at weak disorder but also, for the system sizes accessible to exact diagonalisation,
at intermediate disorder strengths within the ergodic phase and sufficiently far from the transition. We will return to a more sophisticated analysis near the critical point in Secs.~\ref{sec:ipr} and \ref{sec:lengthscale}.

In the MBL phase, the data in Figs.~\ref{fig:Fr} and \ref{fig:r-rsq} are again consistent with a binomial form, $F(r)=\binom{L}{r}p^r(1-p)^{L-r}$, but with $p(W)<1/2$; which implies
\eq{
	\frac{\overline{\braket{r}}_\mathrm{MBL}}{L}=p(W);\quad\frac{\overline{\braket{(\delta r)^2}}_\mathrm{MBL}}{L} = p(W)[1-p(W)]\,,
	\label{eq:r-MBL}
}
and with $p(W)$ decreasing monotonically with $W$. This is well-evidenced in the data in Fig.~\ref{fig:r-rsq}, where both $\overline{\braket{r}}/L$ and $\overline{\braket{(\delta r)^2}}/L$ are $L$-independent in the MBL phase. 

Deep in the MBL phase, the model Eq.~\ref{eq:ham} is perturbatively connected to the non-interacting limit of $J_i = 0$ (referred to as $\mathrm{MBL}_0$ henceforth). Here, while the system is `trivially' MBL since it is a set of non-interacting spins,  the behaviour on the Fock-space is nevertheless non-trivial. The correlation function can be obtained exactly in this limit (see Appendix~\ref{app:mbl0Fr} for details),
\eq{
	\overline{F_{\mathrm{MBL}_0}(r)}=\binom{L}{r}p^r (1-p)^{L-r};\quad p=\frac{\tan^{-1}(W)}{2W}\,,
	\label{eq:F-MBL*}
}
and  the corresponding $\overline{\braket{r}}_{\mathrm{MBL}_0}$ and $\overline{\braket{(\delta r)^2}}_{\mathrm{MBL}_0}$  are shown by the black dashed lines in Fig.~\ref{fig:r-rsq}. Note that the effective $p(W)$ is slightly greater in the interacting MBL phase compared to the non-interacting one Eq.~\ref{eq:F-MBL*},  reflecting the fact that eigenstates of the former are in relative terms less localised on the Fock-space due to the presence of interactions. From Fig.~\ref{fig:r-rsq},  as well as Fig.~\ref{fig:M}, it is also evident that the $\mathrm{MBL}_0$ limit is indeed approached asymptotically with increasing disorder strength $W$.

The picture  suggested by the data is then, inside either phase, that $\overline{F(r)}$ is described by a binomial distribution $B(L,p)$, with $p=1/2$ in the ergodic phase and $p<1/2$ in the MBL phase. As the MBL transition is approached from the localised side, $p(W)$ appears to 
approach a value strictly less than 1/2, suggesting that $\overline{\braket{r}}/L$  -- and consequently $\M$ -- is discontinuous across the transition. While the behaviour near the critical point is not completely clear from the data shown here, it will be substantiated  in 
Secs.~\ref{sec:ipr} and \ref{sec:lengthscale}  via a more sophisticated analysis. It is also seen from the data, Fig.~\ref{fig:r-rsq}(b), that $\braket{(\delta r)^2}/L$ appears to grow unboundedly with $L$ at the critical point. In Sec.~\ref{sec:xiFdist} we will in fact argue that this divergence is of the form $\braket{(\delta r)^2}/L\sim L$.

\begin{figure}
\includegraphics[width=\linewidth]{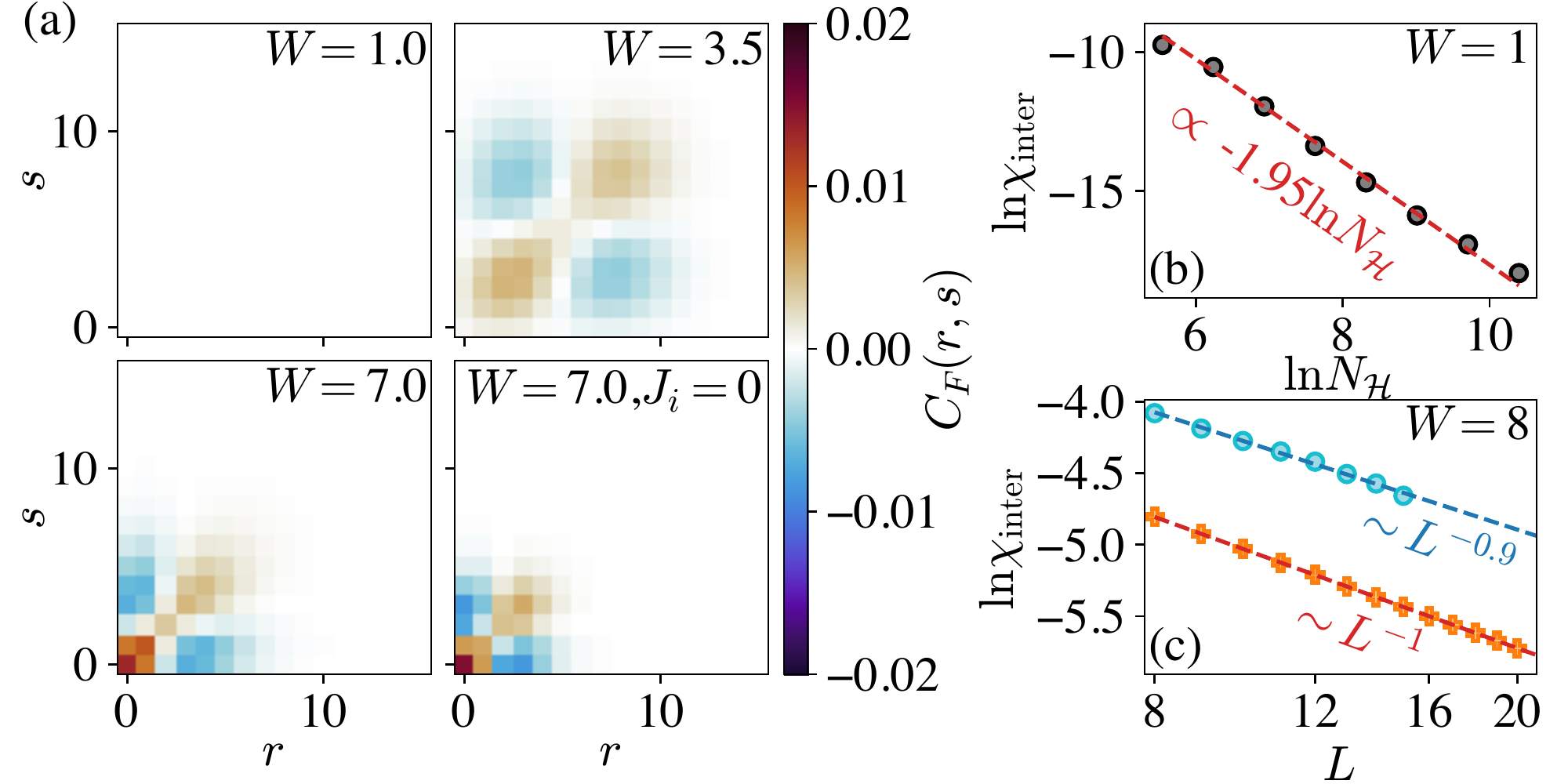}
\caption{
(a) Correlation function $C_F(r,s)$ (Eq.~\ref{eq:Cf}), represented as a colour-map in the $(r,s)$ plane for different disorder strengths, with 
$L=15$. Lower-right panel shows results for the non-interacting limit. (b) In the ergodic phase, the red dashed line shows a fit to the form,  $\ln\inter = -a \ln \nh + b\ln\ln\nh$ with $a=1.95$, implying that $\inter$ decays exponentially with $L$.  (c) In the MBL phase, $\inter$ decays as a power law in $L$ with an exponent $\sim 1$.  Blue data corresponds to the interacting model, and orange data to the non-interacting limit.}
\label{fig:CFrs}
\end{figure}

We turn next to fluctuations of the local polarisation, and discuss how their behaviour is manifest in the Fock-space correlations. In particular we focus on $\inter$, defined in Eq.~\ref{eq:chi-inter}, as it exhibits a peak at the MBL transition (Fig.~\ref{fig:chi}(b)). Using the relation Eq.~\ref{eq:rIK} between the spin-configurations on the Fock-space sites and the distances between them, together with the definiton Eq.~\ref{eq:Fr} of $F(r)$, $\inter$ can be expressed as
\eq{
	\inter = \frac{4}{L^2}\sum_{r,s=0}^L r s~C_F^{\pd}(r,s)\,,\quad\mathrm{where}\nonumber\\
	C_F^{\pd}(r,s) = \left[\overline{F(r)F(s)}-\overline{F(r)}~\overline{F(s)}\right]\,.\label{eq:Cf}
}
Physically, $C_F(r,s)$ encodes the fluctuation over disorder realisations of how the eigenstate is distributed over Fock-space sites distant by $r$ and $s$ simultaneously. Consider for simplicity $r=s$, where $C_F(r,r)=\overline{F^2(r)}-\overline{F(r)}^2$ is simply the fluctuation over disorder realisations of the correlation function $F(r)$. It takes a finite (positive) value in a case where there is a strong inhomogeneity in the spread of the eigenstate on the Fock space. However, the sum rule $\sum_{r,s}C_F(r,s)=0$ then  requires that $C_F(r,s)<0$ for some $r\neq s$. Such a pattern is indeed visible near the MBL transition and in the MBL phase in the numerical results shown in Fig.~\ref{fig:CFrs}(a), indicating the inhomogenous nature of the eigenstate on Fock space in that regime. The structure in $C_F(r,s)$ characteristic of the MBL phase decays with increasing $r$ and $s$ in such a fashion that $\inter$ (related to $C_F(r,s)$ via Eq.~\ref{eq:Cf}) decays with $L$ as a power-law, as is shown explicitly in Fig.~\ref{fig:CFrs}(c); and with increasing $W$, the  structure is increasingly well captured by that arising in the $\mathrm{MBL}_0$ limit, also shown in  Fig.~\ref{fig:CFrs}(a) for $W=7$.

In the ergodic phase at weak disorder strength by contrast, using the fact that the $A_I\sim\mathcal{N}(0,\nh^{-1})$ it can be shown that  
$|C_F(r,s)|\sim \nh^{-2}$.  It is thus completely featureless in the thermodynamic limit, indicating the homgeneity of the spatial correlations, as also seen clearly in Fig.~\ref{fig:CFrs}(a) for $W=1$. This also suggests that $\inter$ should decay to zero exponentially with $L$ in the ergodic phase, evidence for which is presented in Fig.~\ref{fig:CFrs}(b); $\inter\sim \nh^{-2}$ modulo $\ln\nh$ corrections.

%%%%%%%%%%%%%%%%%%%%%%%%%%%%%%%%%%%%%%%%%%%%%%%%%%%%%%%%%%%%%%%%%%%%%%%%%%%%%%%%%

\subsection{Radial probability distribution \label{sec:radialprob}}
A quantity related to $F(r)$ was studied extensively in Ref.~\cite{detomasi2020rare}. Referred to as the radial probability distribution, it is defined for an eigenstate $\ket{E}=\sum_{I}A_I\ket{I}$ as 
\eq{
	\Pi(r) = \sum_{K: r_{I_0K}^{\pd}=r}|A_K^{\pd}|^2\,,
	\label{eq:Pir}
}
where $I_0$  is the Fock-space site on which the wavefunction has maximum amplitude, i.e.\ $|A_{I_0}|^2=\max_I|A_I|^2$. 
Physically, $\Pi(r)$ is a measure of the relative support of the eigenstate on Fock-space sites at Hamming distance $r$ from $I_0$, the Fock-space site on which the state is localised for $\Gamma=0$. The connection between $F(r)$ and $\Pi(r)$ follows by noting that $F(r)$  (Eq.~\ref{eq:Fr}) can be written as $F(r)=\sum_{I}|A_I|^2 F_I(r)$ where $F_I(r)=\sum_{K: r_{IK}=r}|A_K|^2$, from which it is obvious that  $\Pi(r)=F_{I_0}(r)$. While $\Pi(r)$ is certainly informative (we briefly consider it further in Sec.~\ref{sec:xiPi}), one advantage of $F(r)$ is that it is a
correlation function summed over all Fock-space sites, and as such does not make any Fock-space site `special'. $\Pi(r)$ also relies on being able to identify the Fock-space site $I_0$, which is naturally ambiguous in the ergodic phase, and potentially so in the MBL phase due to the fractal scaling of eigenstate amplitudes.

Since $\sum_{r=0}^L\Pi(r)=1$ due to wavefunction normalisation, $\Pi(r)$, like $F(r)$, can be interpreted as a probability distribution on the Fock space, such that one can define a mean distance and an associated variance as
\eq{
	\braket{d}=\sum_{r=0}^L r\Pi(r)\,,\quad\braket{(\delta d)^2}=\sum_{r=0}^L r^2\Pi(r)-\braket{d}^2\,.
	\label{eq:d-dsq}
}
The form of $\Pi(r)$ bears many similarities to that of $F(r)$. We refer to Ref.~\cite{detomasi2020rare} for a detailed analysis of $\Pi(r)$ and briefly summarise the results here.  In the ergodic phase, since the wavefunctions are uniformly spread on the Fock space, $\Pi(r)=2^{-L}\binom{L}{r}$ with $\overline{\braket{d}}=L/2$ and $\overline{\braket{(\delta d)^2}}=L/4$; as expected, this is the same behaviour as arises for $F(r)$. In the MBL phase $\Pi(r)$, like $F(r)$, is described by a binomial distribution $\Pi(r)=\binom{L}{r}[p_\Pi(W)]^r[1-p_\Pi(W)]^{L-r}$ with $p_\Pi(W)<1/2$. This is qualitatively similar to the results for $F(r)$ (Eq.~\ref{eq:r-MBL}), although  $p_\Pi(W)$ differs quantitatively from $p(W)$. In particular, in the $\mathrm{MBL}_0$ case, $p_\Pi(W)= [W-(\sqrt{W^2+1}-1)]/2W$~\cite{detomasi2020rare}, which 
differs from the corresponding quantity $p(W)$ for $F(r)$ in Eq.~\ref{eq:F-MBL*}. Finally, at the critical point, $\Pi(r)$ is broad in $r$ such that $\overline{\braket{d}}=p_\Pi(W_c)L$ where $p_\Pi(W_c)<1/2$ but the variance scales as $\overline{\braket{(\delta d)^2}}\sim L^a$ with $a>1$.

%%%%%%%%%%%%%%%%%%%%%%%%%%%%%%%%%%%%%%%%%%%%%%%%%%%%%%%%%%%%%%%%%%%%%%%%%%%%%%%%%%%%%%%%%%%%%%%%%%%%%

\section{Eigenstate correlations and Fock-space IPR \label{sec:ipr}}

The distribution of wavefunction amplitudes on the Fock space is central in characterising eigenstates in either phase. They are often quantified via generalised participation entropies (PE) or inverse participation ratios (IPR). The $q^\mathrm{th}$ IPR for an eigenstate $\ket{E}=\sum_{I}A_I\ket{I}$ is defined as 
\eq{
\mathcal{L}_q^{\pd}=\sum_{I}|A_I|^{2q} = \Lambda_q^{\pd} \nh^{-\tau_q}\,,
\label{eq:Lq}
} 
where $\tau_q$  defines the fractal dimension. For an ergodic phase $\tau_{q}=(q-1)$, whereas an MBL phase is signalled by $\tau_{q}< (q-1)$.  
As elaborated below, in the ergodic phase the multiplicative factor $\Lambda_q >1$  can be understood as a non-ergodicity volume, while $\Lambda_q <1$ throughout the MBL phase and $\Lambda_q=1$ at the MBL transition~\cite{mace2019multifractal}. In the following, we will focus on the case of $q=2$, since $\ltwo$ is directly related to the Fock-space correlation $F(r)$ (Eq.~\ref{eq:Fr}) by 
\eq{
	F(r=0) = \sum_I |A_I|^4 = \ltwo\,.
	\label{eq:L2}
}
A scaling theory of the MBL transition based on the IPRs in Fock-space is clearly desirable, for it sheds light on the critical behaviour of local observables such as eigenstate polarisations, via their relation to the Fock-space correlations (see Eqs.~\ref{eq:rmean},\ref{eq:Ms-r}).

In a comprehensive analysis~\cite{mace2019multifractal}, Mac\'e \textit{et al.} showed that the MBL transition in a disordered XXZ chain can be described by a scaling theory for the PEs, in which the scaling forms are asymmetrical on the two sides of the transition: on the ergodic side the PEs follow a `volumic' scaling form,  while on the MBL side, they follow a  `linear' scaling form.  Here we perform a similar analysis for the disordered Ising chain Eq.~\ref{eq:ham}, focusing on the IPRs. The results arising form an important element (Sec.~\ref{sec:lengthscale} ff) in understanding the behaviour of emergent lengthscales in Fock space, in either phase as well as their critical behaviour.
Following Ref.~\cite{mace2019multifractal}, we use the scaling ansatz
\eq{
	\ipr/\overline{\mathcal{L}_{2,c}} = \begin{cases}
										\mathcal{F}_\mathrm{vol}^{\pd}\left(\frac{\nh}{\Lambda_2}\right)~&:~ W<W_c\\
										\mathcal{F}_\mathrm{lin}^{\pd}\left(\frac{\ln\nh}{\xi}\right)~&:~ W>W_c
										\end{cases}\,,
\label{eq:ipr-scaling}
}
where $\overline{\mathcal{L}_{2,c}}=\nh^{-\tau_{2,c}}$ is the average IPR as the critical point is approached from the MBL side, 
$W\to W_{c}^{+}$. Such an ansatz is rooted in the scaling theory of Anderson transitions on random graphs formally in infinite-dimension~\cite{garciamata2017scaling}.

\begin{figure}
\includegraphics[width=\linewidth]{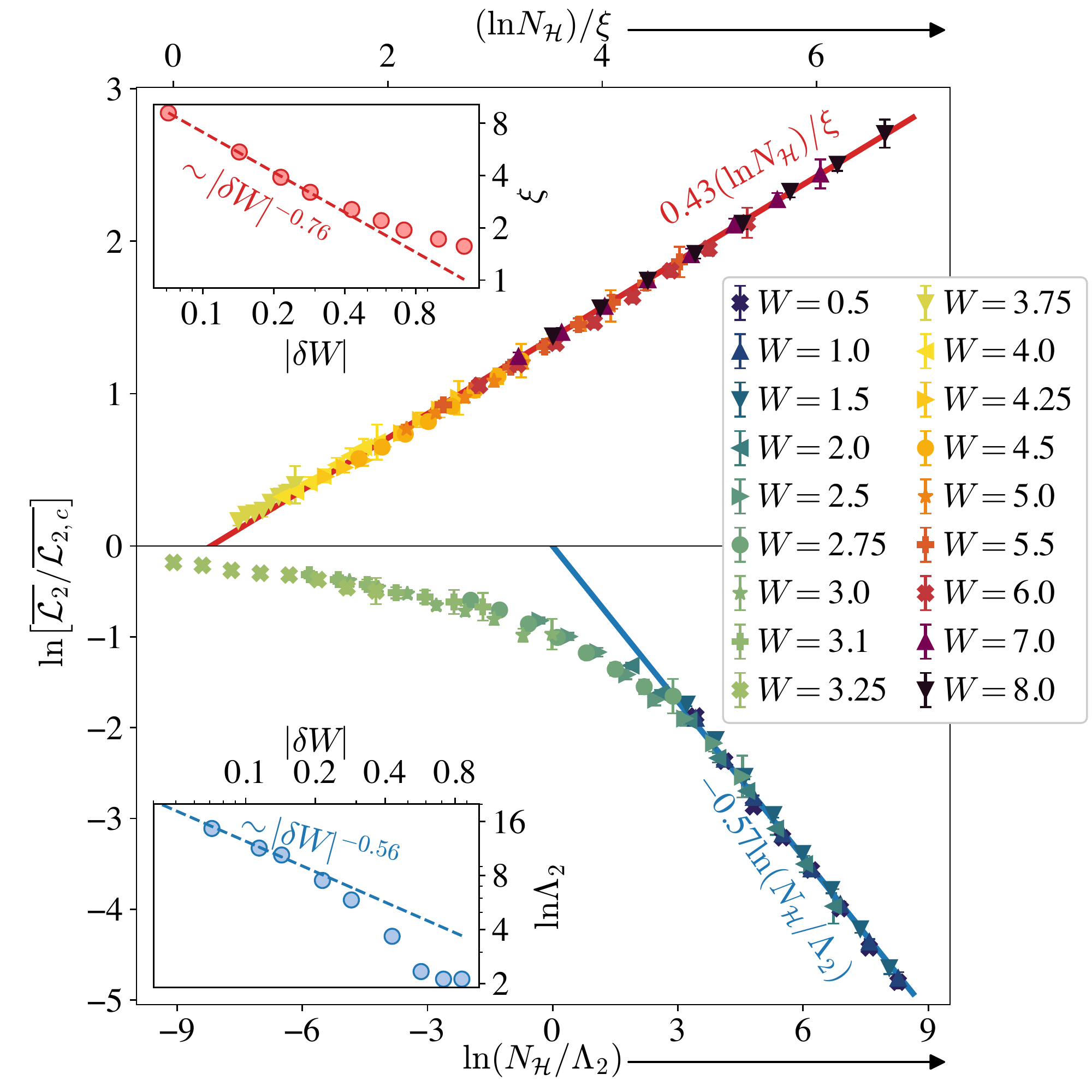}
\caption{
Scaling of the mean IPR, $\ipr$, relative to its critical value, $\overline{\mathcal{L}}_{2,c}$:
$\ln[\ipr/\overline{\mathcal{L}}_{2,c}]$ \emph{vs} $(\ln \nh)/\xi$ for the MBL phase (top panel), and \emph{vs}
$\ln[\nh/\Lambda_{2}]$ for the ergodic phase (lower panel). MBL phase data shows collapse onto a scaling function of $(\ln\nh)/\xi$, where the lengthscale $\xi$ diverges $\xi \sim (W-W_c)^{-\beta}$ with $\beta\approx 0.76$ (upper inset). Ergodic phase data collapses onto a scaling function of $\nh/\Lambda_2$, where the volume $\Lambda_2$ diverges $\Lambda_2 \sim \exp([W_c-W]^{-\alpha})$ with $\alpha\approx0.5$ (lower inset). The analysis used $W_c=3.5$, with data obtained for system sizes $L=8$$-$$15$.
}
\label{fig:ipr-scaling}
\end{figure}

In the ergodic phase $\tau_2=1$ is expected asymptotically ($L\to \infty$), such that $\ipr = \Lambda_2/\nh$ with $\Lambda_2$ interpreted as a non-ergodicity volume~\cite{mace2019multifractal}. For scales $\nh \ll \Lambda_{2}$, one's physical intuition is that eigenstates reside sparsely on the Fock space, while for scales $\nh \gg \Lambda_{2}$ the sparse structure repeats itself and leads to full ergodicity on the largest scales. This motivates a volumic scaling function $\mathcal{F}_\mathrm{vol}(\nh/\Lambda_2)$ of form
\eq{
	\mathcal{F}_\mathrm{vol}^{\pd}(x)= \begin{cases}
								x^{(\tau_{2,c}^{\pd}-1)}~ & :~ x\gg 1\\
								1~ &:~ x\to 0
								\end{cases}\,.
\label{eq:Fvol}
}
From this scaling function for $\ipr/\overline{\mathcal{L}_{2,c}}$, one obtains $\ipr=\Lambda_2^{1-\tau_{2,c}}/\nh$ in the asymptotic limit $\nh\gg\Lambda_2$, while in the opposite limit $\nh\ll\Lambda_2$ one has the critical scaling $\ipr\to\overline{\mathcal{L}_{2,c}}$.  As 
evident in Fig.~\ref{fig:ipr-scaling} (lower panel), a volumic scaling function indeed leads to an excellent collapse of the  numerical
data in the ergodic phase, with the non-ergodicity volume found to diverge at the critical point as $\ln\Lambda_2\sim |\delta W|^{-\alpha}$ 
with a critical exponent $\alpha\approx 0.5$, where $\delta W = (W-W_c)/W_c$. The asymptotic form of the numerically determined scaling function is moreover seen to be in excellent agreement with  Eq.~\ref{eq:Fvol} with $\tau_{2,c}\approx 0.43$.

Turning now to the MBL phase, the physical intuition here is that eigenstates are supported sparsely on some `strands' on the Fock-space graph
with a finite correlation length $\xi_{F}$ on the strands. This motivates the linear scaling function~\cite{mace2019multifractal} 
$\mathcal{F}_\mathrm{lin}(\ln(\nh)/\xi)$ for $\ipr/\overline{\mathcal{L}_{2,c}}$,  with an emergent lengthscale $\xi$. In the limit  
$\ln\nh\propto L \ll  \xi$ the critical scaling $\ipr\to\overline{\mathcal{L}_{2,c}}$ arises, whereas in the opposite limit $\ln\nh \propto L \gg\xi$ one expects $\ipr\sim \nh^{-\tau_2}\equiv\nh^{-\tau_{2,c}}\mathcal{F}_\mathrm{lin}(\ln(\nh)/\xi)$. This suggests the scaling function to have the asymptotic behaviour
 \eq{
	\mathcal{F}_\mathrm{lin}^{\pd}(x)= \begin{cases}
								e^{\tau_{2,c}x}~ & :~ x\gg 1\\
								1~ &:~ x\to 0
								\end{cases}\,
\label{eq:Flin}
}
(with $x=(\ln\nh)/\xi$), where
\eq{
	\xi = (1-\tau_2^{\pd}/\tau_{2,c}^{\pd})^{-1}\,
	\label{eq:xi-tau2}
}
provides a direct relation between the lengthscale $\xi$ and the IPR fractal dimensions (and we will further relate $\xi$ to $\xi_{F}$ 
in Sec.~\ref{sec:xiF}).

The numerical results of Fig.~\ref{fig:ipr-scaling} (upper panel) indeed show that with the linear scaling ansatz $\mathcal{F}_\mathrm{lin}((\ln\nh)/\xi)$, the data exhibits excellent scaling collapse, with $\xi\sim |\delta W|^{-\beta}$ found to diverge on approaching the transition, 
with exponent $\beta\approx 0.8$. The collapsed data is also in excellent agreement with the aysmptotic form of the scaling function in Eq.~\ref{eq:Flin} with $\tau_{2,c}\approx 0.43$; which, consistently, is the same value obtained from the scaling analysis on the ergodic side.

As with the disordered XXZ chain~\cite{mace2019multifractal}, the picture that emerges is thus as follows.  The entire ergodic phase is characterised by $\tau_2=1$, with a non-ergodicity volume $\Lambda_2$ which diverges on approaching the transition from the ergodic side as $\Lambda_2\sim e^{(W_c-W)^{-\alpha}}$ with $\alpha\approx 0.5$.  As the MBL transition is crossed, $\tau_2$ jumps discontinuously to a value of $\tau_{2,c}\approx 0.43$ for the disordered spin-1/2 chain Eq.\ \ref{eq:ham} in the $\sigma^z$-basis, indicating that the eigenstates at the critical point have fractal statistics. 

 In fact, the entire MBL phase hosts fractal eigenstates on the Fock space, with $\tau_2$ decreasing continuously from $\tau_{2,c}$
upon increasing the disorder strength above $W_{c}$. This naturally suggests the critical point belongs to the same non-ergodic class as the rest of the MBL phase, and hence that it can be  regarded as the terminal end-point of a line of fixed points (in the RG sense) characterising the MBL phase. This interpretation, along with the form of the divergence of the non-ergodic volume at the transition on approaching it from the ergodic side, is consistent with a Kosterlitz-Thouless-like scaling at the MBL transition~\cite{thiery2018many-body,dumitrescu2018kosterlitz,morningstar2020manybody,laflorencie2020chain,hopjan2021detecting}.

In the analysis above, the exponent $\tau_2$ was that obtained from the average IPR, $\ipr$. However, as evident from Eq.~\ref{eq:Lq} for $\mathcal{L}_{2}$,  in general $\tau_2$ has a distribution over disorder realisations, and the discontinuous jump in $\tau_2$ across the MBL transition naturally invites questions about the distribution. To obtain further insight into this, we extract the distribution $P_{\tau_2}$ numerically by noting (Eq.~\ref{eq:Lq}) that
\eq{
\tau_2^{\pd} = -\frac{\ln\ltwo}{\ln\nh}+\frac{\ln\Lambda_2^{\pd}}{\ln\nh}\,, 
\label{eq:tau2-dist}
}
and transforming over the distributions of $\mathcal{L}_2$. The distribution of $\tau_2$ was also studied in Ref.~\cite{detomasi2020rare}, where the contribution from the second term in Eq.~\ref{eq:tau2-dist} was neglected. While the latter is certainly valid in the thermodynamic limit, we find  that for the system sizes accessible to exact diagonalisation this contribution is not in fact negligible. To take it into account, we thus extract it as the intercept to a linear fit of $\ln\ipr$ versus $\ln\nh$  (taking $\Lambda_2$ to be $\delta$-distributed).

\begin{figure}
\includegraphics[width=\linewidth]{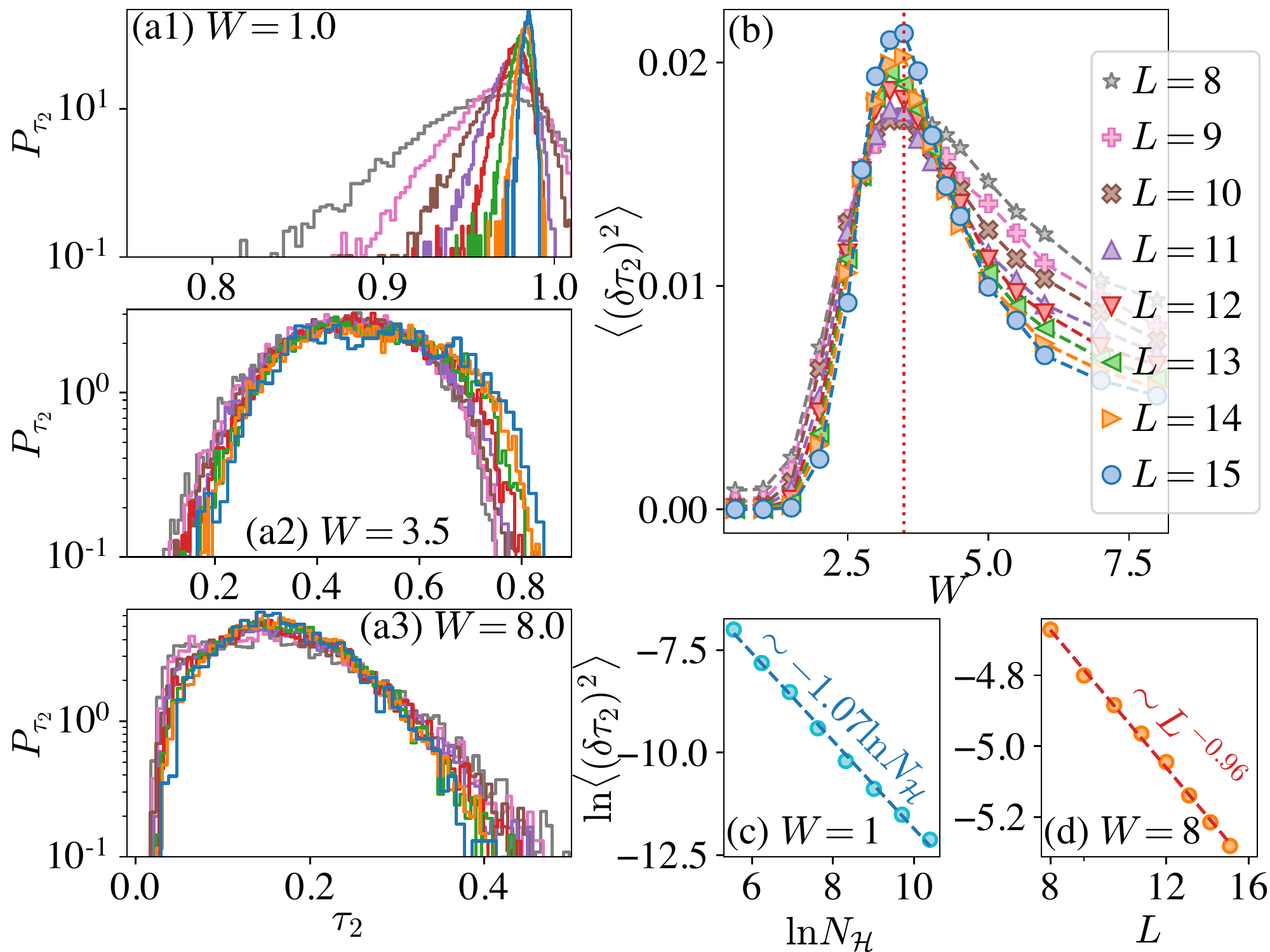}
\caption{Distributions of IPR fractal exponent, $\tau_2$, across the MBL transition. In the ergodic phase [panel (a1)], $P_{\tau_2}$ rapidly converges to a $\delta$-function at $\tau_2=1$. In the MBL phase  [panel (a3)], $P_{\tau_2}$ also narrows with increasing $L$, with a peak at $\tau_2<1$ indicating the fractal nature of  MBL eigenstates. This is reflected in the variance of $\tau_2$ decreasing with $L$ in either phase; in the ergodic phase it decays exponentially with $L$ [panel (c)], while in the MBL phase it decays $\sim 1/L$ [panel (d)].
At the critical point by contrast [panel (a2)], the distribution does not narrow with increasing $L$, and the variance appears to saturate with $L$ to a finite value [panel (b)].
}
\label{fig:tau-dist}
\end{figure}

The results for $P_{\tau_2}$ are shown in Fig.~\ref{fig:tau-dist}. In the ergodic phase (Fig.~\ref{fig:tau-dist}(a1)), 
the distribution rapidly converges towards a $\delta$-function at $\tau_2=1$. This is confirmed by the variance of the distribution, $\braket{(\delta \tau_2)^2}$ (panel (c)), vanishing $\sim \nh^{-1}$ exponentially with $L$. Deep in the ergodic phase, this can be understood simply by 
regarding the eigenstate amplitudes to be normally distributed independent random numbers, with zero mean and a variance $\nh^{-1}$. In the MBL phase as well (Fig.~\ref{fig:tau-dist}(a3)), the distribution $P_{\tau_2}$ narrows with increasing $L$. Here however it does so in a qualitatively slower fashion compared to the ergodic phase, with the variance  $\braket{(\delta \tau_2)^2} \propto 1/L$ as shown numerically in Fig.~\ref{fig:tau-dist}(d). In the $\mbln$ case, it can in fact be shown analytically that $\braket{(\delta \tau_2)^2}\sim 1/L$ (see Appendix~\ref{app:mbl0ipr}).  At the critical point, the situation is qualitatively different from either of the two phases: the distribution $P_{\tau_2}$ (Fig.~\ref{fig:tau-dist}(a2)) is broad in the sense that the variance $\braket{(\delta \tau_2)^2}$ (panel (b)) remains finite 
as $L\to \infty$ ($P_{\tau_2}$ has of course a strictly bounded support $\in [0,1]$, whence none of its moments can diverge).

%%%%%%%%%%%%%%%%%%%%%%%%%%%%%%%%%%%%%%%%%%%%%%%%%%%%%%%%%%%%%%%%%%%%%%%%%%%%%%%%%%%%%%%%%%%%%%%%

\section{Lengthscales on Fock-space \label{sec:lengthscale}}
We turn now to the central question of how a lengthscale emerges on the Fock space from the spatial correlations embodied in 
$\overline{F(r)}$. Based on the relation between $\Fr$ and the IPR $\ipr$, together with a scaling theory for the latter (Sec.~\ref{sec:ipr}), we infer the behaviour of this lengthscale in each phase, as well as in the vicinity of the transition. The critical behaviour of $\xi_F$ can then be used to obtain the critical behaviour of local observables, such as local polarisations, by exploiting the relation between them and the Fock-space correlation function.

%%%%%%%%%%%%%%%%%%%%%%%%

\subsection{Correlation length \label{sec:xiF}}

The Fock-space correlation function $F(r)$ (Eq.~\ref{eq:Fr}) entails a sum of the eigenstate amplitudes $|A_I|^2|A_K|^2$, over all pairs of Fock-space sites separated by a given Hamming distance $r_{IK}=r$. Sufficiently deep in an ergodic phase, one expects $|A_I|^2|A_K|^2$ to be independent of $r$ (and $\sim\nh^{-2}$). In an MBL phase by contrast, our physical intuition is that the eigenstate correlation $|A_I|^2|A_K|^2$ will, on an average, decay exponentially with $r_{IK}$. At the same time, however, the number of Fock-space sites at distance $r$ from any given Fock-space site, $N_r =\binom{L}{r}$, grows exponentially with $r$ (for $r\le L/2$).  As we will show shortly, it is the competition between these exponentials in the correlation function, together with the normalisation of $F(r)$,  which lies at the heart of the fractal scaling of the IPR in the MBL phase.

From the above considerations, one anticipates the disorder averaged correlation function $\Fr \propto N_r e^{-r/\xi_F}$, where $\xi_F$ denotes
the lengthscale for exponential decay of the average correlation function between any two Fock-space sites. Imposing the normalisation $\sum_{r=0}^{L} F(r)=1$ (as follows exactly from the definition Eq.~\ref{eq:Fr}), gives 
\eq{\Fr = N_r\frac{e^{-r/\xi_F}}{(1+e^{-1/\xi_F})^{L}}\,; 
\label{eq:Frxi}
}
which can be recast equivalently as 
\eq{
	\Fr \equiv \nh N_r^{\pd} f(r);\quad f(r)=\frac{e^{-r/\xi_F}}{2^L(1+e^{-1/\xi_F})^{L}} \,,
	\label{eq:xiF-def}
}
where $f(r)$ encodes the average eigenstate correlation between two Fock-space sites a Hamming distance $r$ apart.

\begin{figure}
\includegraphics[width=\columnwidth]{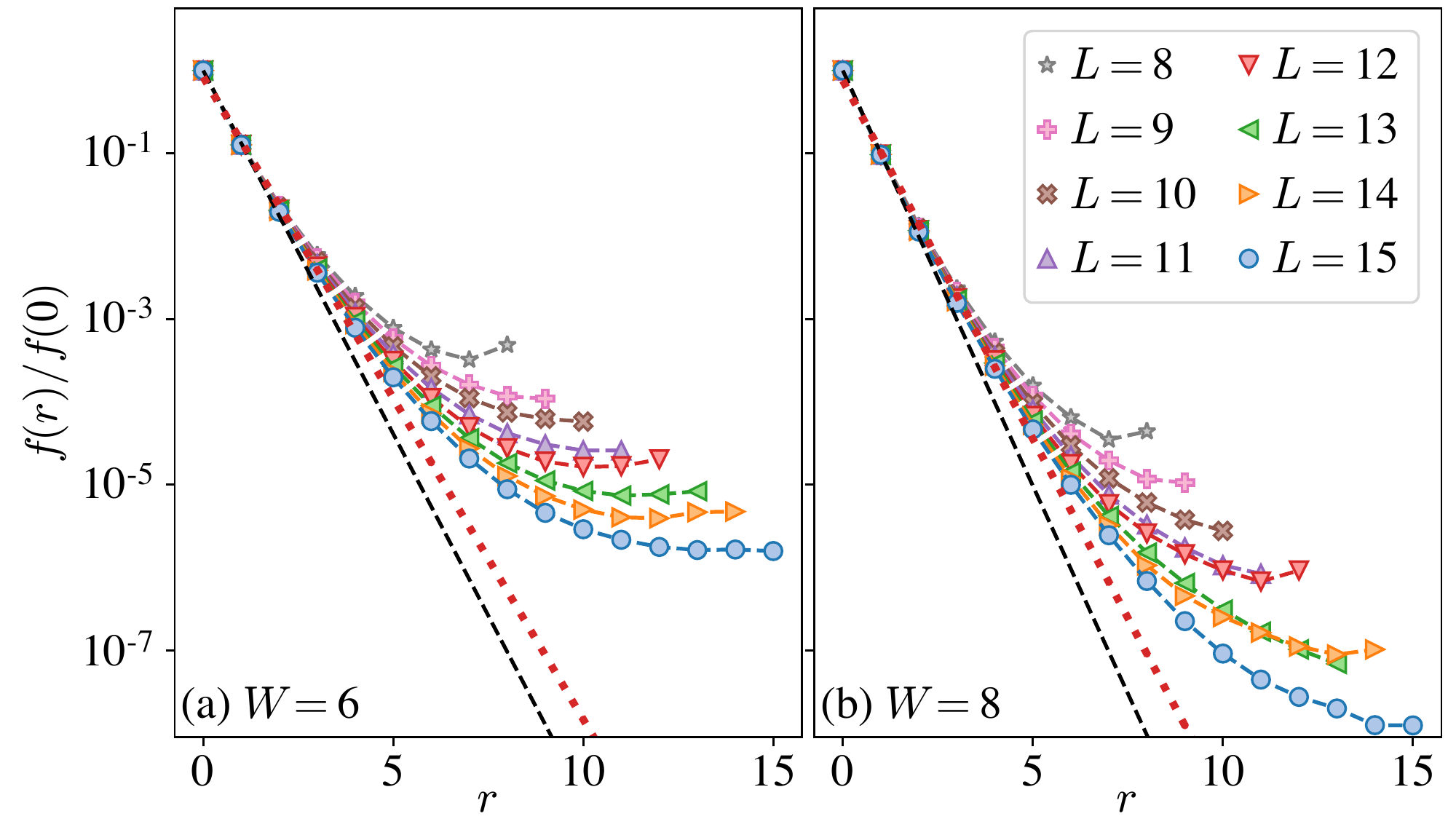}
\caption{Exponential decay in the MBL phase of the rescaled correlation function, $f(r)$ (Eq.~\ref{eq:xiF-def}), with a finite correlation length $\xi_{F}$. With increasing system size, data falls on the exponential decay (indicated by red dotted lines) for a larger range of $r$. 
Black dashed lines show the $\mbln$ case, which naturally has a slightly smaller correlation length.
}
\label{fig:frf0}
\end{figure}

Fig.~\ref{fig:frf0} presents numerical results for $f(r)$, for two values of $W$ in the MBL phase. These indeed show clearly the exponential decay of $f(r)$, with an $L$-independent lengthscale. With increasing system size for given $W$, $f(r)$ falls onto the common exponential decay for a larger range of $r$. 

While the arguments above for the form of $\overline{F(r)}$ were physically motivated, it is important to note that Eq.~\ref{eq:xiF-def} also arises constructively in the `strong disorder' $\mbln$ case of $J_{i}=0$ (which is perturbatively connected to the interacting MBL phase for sufficiently large $W$).  Here, from Eq.~\ref{eq:F-MBL*}, one has directly that
 \eq{
	f_{\mbln}^{\pd}(r) = [2/(1-p)]^{-L}e^{-r\ln[(1-p)/p]}\,.
}
This is precisely of form Eq.~\ref{eq:xiF-def}, with the correlation length given by
\eq{
	\xi_F^{-1} = \ln\big[(1-p)/p\big]~~~~:\mbln\,,
}
and $p \equiv p(W)$ given explicitly in Eq.~\ref{eq:F-MBL*}. Note that, for a given disorder strength, $\xi_F$ for the $\mbln$ case 
(also indicated in  Fig.~\ref{fig:frf0}) is slightly smaller than that for the interacting MBL phase, as the latter is naturally less localised than the former.

Having established the presence of a lengthscale, $\xi_F$, on the Fock space via the correlation function $\overline{F(r)}$, we can exploit the latter's relation to the IPR to study the properties of $\xi_F$ across the MBL transition. The central relation here, from Eq.~\ref{eq:L2} and Eq.~\ref{eq:xiF-def}, is 
\eq{
	\ln(\ipr) = -L\ln\left(1+e^{-1/\xi_F}\right)\,.
	\label{eq:L2xiF}
}
Recalling (Sec.~\ref{sec:ipr}) that $\ln(\ipr) = -L\times \tau_2\ln2$ in the asymptotic limit in either phase, Eq.~\ref{eq:L2xiF} leads to the identification 
\eq{
	\tau_2^{\pd} = \ln\left(1+e^{-1/\xi_F}\right)/\ln 2\,.
	\label{eq:tau2xiF}
}
A finite $\xi_F$ in Eq.~\ref{eq:tau2xiF} implies that $\tau_2<1$, which shows how the fractal nature of  MBL eigenstates on the Fock-space emerges out of the competition between an exponential decay of $f(r)$ and the exponential growth of $N_{r} =\binom{L}{r}$.
As the fractality of the eigenstates persists throughout the MBL phase, the correlation length $\xi_F$ is finite in the entire phase. Throughout the ergodic phase by contrast, $\tau_2=1$, which is consistent with a divergent $\xi_F$ in the asymptotic limit. 
Indeed, directly setting $\xi_{F}=\infty$ in Eq.~\ref{eq:Frxi} gives $\overline{F(r)}=2^{-L}N_{r}$, which is just the behaviour
observed numerically (Fig.~\ref{fig:Fr}) deep in the ergodic phase.

With the behaviour of $\xi_F$ established in the two phases asymptotically, we turn next to their behaviour at and in the vicinity of the critical point. First, recall (Sec.~\ref{sec:ipr}) that at the MBL transition one has $\ln\overline{\mathcal{L}_{2,c}}=-L\tau_{2,c}\ln 2$ which, using Eq.~\ref{eq:L2xiF}, gives 
\eq{
	\xi_{F,c}^{-1}=-\ln(2^{\tau_{2,c}}-1)\,.
}
Since ($0<$) $\tau_{2,c}<1$, this implies that the correlation length at the MBL transition is finite; which is consistent with the phenomenology that the eigenstates at the critical point lie in the same class of non-ergodic multifractal states as the entire MBL phase.

The critical properties of $\xi_F$ on approaching the transition from the MBL side then follow readily. From the scaling behaviour of $\ipr$ in the MBL phase obtained in Sec.~\ref{sec:ipr} and Eq.~\ref{eq:L2xiF}, one obtains
\eq{
	-L\ln\left(1+e^{-1/\xi_F}\right)&=-L\tau_{2,c}^{\pd}(1-1/\xi)\ln 2\nonumber\\
	\Rightarrow~~ \frac{1}{\xi_F}-\frac{1}{\xi_{F,c}}~\overset{\xi\gg 1}{\sim}~&\frac{1}{\xi}\left(\frac{\tau_{2,c}^{\pd}}{1-2^{-\tau_{2,c}}}\ln2\right)\,;
}
thereby relating $\xi_{F}$ to the lengthscale $\xi$ which (Sec.~\ref{sec:ipr}) diverges as $\xi \sim |\delta W|^{-\beta}$ on approaching the transition. This implies that $\xi_{F,c}-\xi_{F}\sim |\delta W|^\beta$ vanishes with the same exponent, as too 
does $\tau_{2,c}-\tau_{2}$ (from Eq.~\ref{eq:xi-tau2}).
\new{We note that similar scaling for the typical localisation length is found for Anderson transitions on random regular graphs~\cite{garciamata2020two},
and for a typical real-space localisation length in phenomenological RG approaches to MBL~\cite{dumitrescu2018kosterlitz}. }

We turn now to critical behaviour on approaching the transition from the ergodic side. Since $\xi_F$ is a lengthscale on the Fock space, 
diverging in the thermodynamic limit throughout the ergodic phase, one naturally expects it to scale as  $\xi_F=\gamma L + b \sim \gamma L$. 
This is concomitant with the volumic scaling of the IPR on the ergodic side of the transition. The scaling $\ipr =\Lambda_{2}\nh^{-1}$ in the ergodic phase gives
\eq{
	-L\ln\left[1+e^{-1/(\gamma L)}\right]&=\ln\Lambda_2^{\pd}-L\ln2\nonumber\\
	\overset{L\gg 1}{\Rightarrow }\gamma &= \frac{1}{2\ln\Lambda_2^{\pd}}\,,
}
which implies that  $\gamma\sim|\delta W|^\alpha$ vanishes as the transition is approached, controlled by the same exponent
$\alpha$ with which the non-ergodicity volume $\Lambda_2\sim e^{(W_c-W)^{-\alpha}}$ diverges (Sec.~\ref{sec:ipr}).

As summarised in Fig.~\ref{fig:summary}, one thus concludes the following. $\xi_{F}$ is finite throughout the MBL phase,  and  tends to a finite  $\xi_{F,c}$ at the critical point on approaching the transition from  the MBL side; $\xi_{F}-\xi_{F,c}$ vanishing as a power-law in $\delta W$, with the same exponent with which $\tau_2$ approaches $\tau_{2,c}$. Across the transition into the ergodic phase, $\xi_F$ jumps discontinously to a divergent value in the thermodynamic limit, consistent with the discontinuous jump in $\tau_2$ to 1. In the ergodic phase, $\xi_F$ diverges as $\gamma L$, where $\gamma$ vanishes at the transition as a power-law in $|\delta W|$, with the same critical exponent with which the logarithm of the non-ergodicity volume, $\ln \Lambda_2$, diverges.

%%%%%%%%%%%%%%%%%%%%%%%%%%%%%%%%%%%%%%%%%%%%%%%%%%%%%%%%

\subsection{Connection to local polarisation \label{sec:local}}

The form of $\overline{F(r)}$ in Eq.~\ref{eq:xiF-def} immediately  yields, via Eq.~\ref{eq:Ms-r}, a direct relation between the length 
scale $\xi_F$ and the average local polarisation $\M$, 
\eq{
	\M=1-2\frac{\overline{\braket{r}}}{L}  =1-2\left(1+e^{1/\xi_F}\right)^{-1}\,.
	\label{eq:Ml-xiF}
}
A finite $\xi_F$ implies $0<\M<1$, as is indeed the case throughout the MBL phase, including at the transition. By contrast,
a divergent $\xi_F$ in the ergodic phase implies $\M=0$ in the thermodynamic limit. This is consistent with the results obtained 
from exact diagonalisation in Sec.~\ref{sec:model-loc-obs}. An important conclusion is therefore that the local polarisation, which is a diagnostic of the MBL transition, is \emph{discontinuous} across the transition, as depicted schematically in Fig.~\ref{fig:summary}. This result is also consistent with predictions based on entanglement entropies, which are spatially global quantities, that the MBL critical point itself is non-ergodic~\cite{khemani2017critical}.

%%%%%%%%%%%%%%%%%%%%%%%%%%%%%%%%%%%%%%%%%%%%%%%%%%%%%%%%%%%%%%%%%%%%%%

\subsection{Distributions of correlation length \label{sec:xiFdist}}

\begin{figure}
\includegraphics[width=\linewidth]{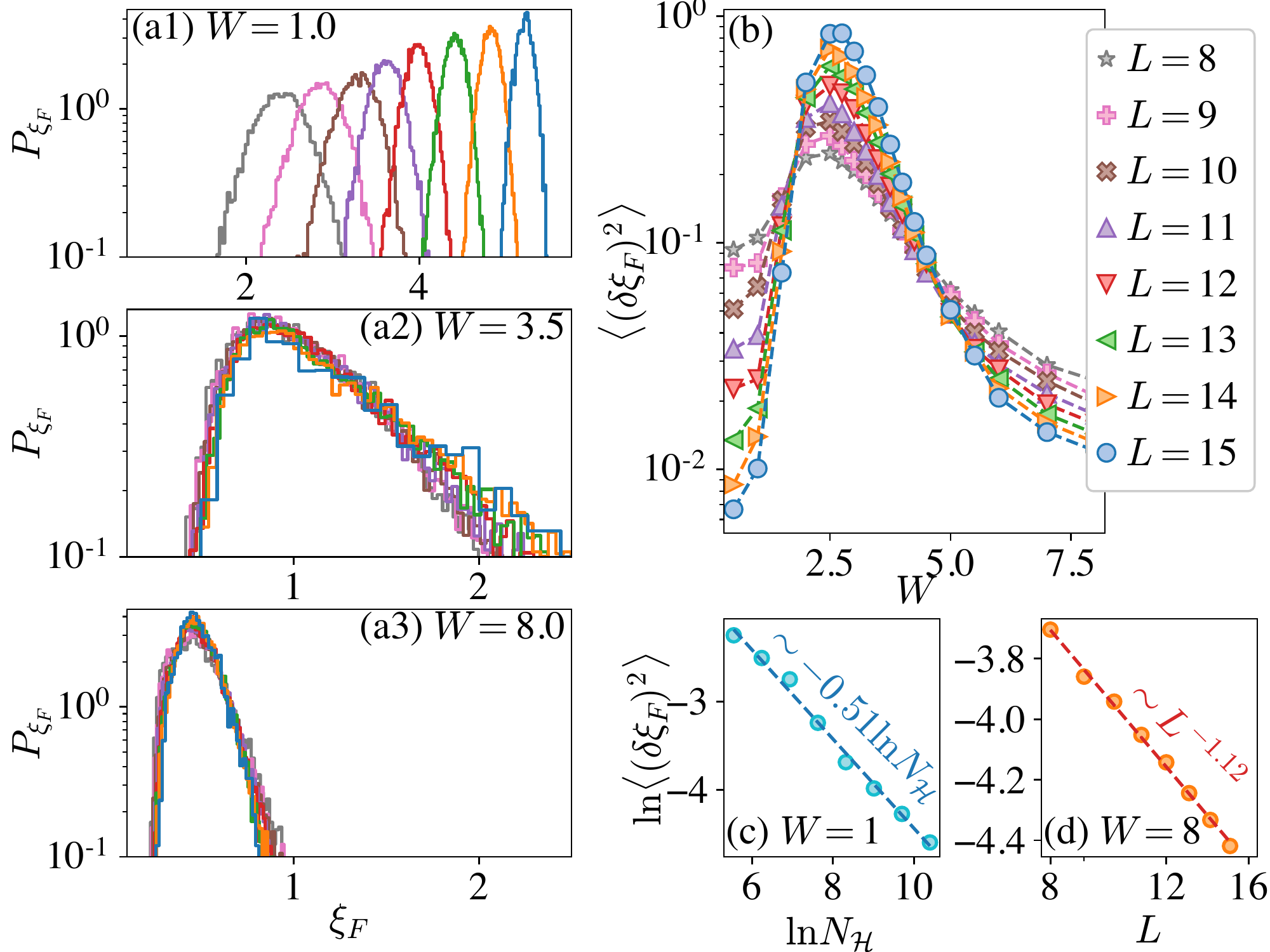}
\caption{Behaviour of the Fock-space correlation length, $\xi_F$, across the MBL transition. Panels (a1)-(a3) respectively show the distributions, $P_{\xi_{F}}$, of $\xi_F$ in the ergodic phase, at the MBL transition, and in the MBL phase. In the ergodic phase, 
$P_{\xi_{F}}$ shifts linearly with $L$ while simultaneoulsy narrowing, implying $\xi_F=\gamma L$ with the distribution of $\gamma$ tending to a $\delta$-function as $L\to\infty$. This is also reflected in the variance $\braket{(\delta \xi_F)^2}$ decaying exponentially with $L$ [panels (b) and (c)]. In the MBL phase $P_{\xi_{F}}$ again narrows with increasing $L$, but its mean converges to an $\mathcal{O}(1)$ value.
In contrast to the ergodic phase, $\braket{(\delta \xi_F)^2}\sim 1/L$ [panel (d)]. At the MBL transition, the distributions visibly broaden with $L$, as reflected in the variance of $\xi_F$ diverging with $L$ [panel (b)].}
\label{fig:xi-dist}
\end{figure}

In the analysis above, it was implicitly assumed that the physics is captured by a single  Fock-space correlation length $\xi_F$ which was obtained from $\ipr$ (Eq.~\ref{eq:L2xiF}); in other words, the distribution $P_{\xi_{F}}(\xi_{F})$ of $\xi_F$ over disorder realisations is 
presumed to tend to a $\delta$-function in the thermodynamic limit, in either phase. To examine this, we study the distribution $P_{\xi_F}$, obtained via $\mathcal{L}_{2}$ from $\mathcal{L}_{2}=(1+e^{-1/\xi_{F}})^{-L}$. As shown in Fig.~\ref{fig:xi-dist}, $P_{\xi_F}$ indeed sharpens with increasing $L$ in each phase. This is quantified by considering the variance of the distribution, $\braket{(\delta \xi_F)^2}$, which decays to zero with $L$ in both phases as shown in Fig.~\ref{fig:xi-dist}(b); with the decay being respectively exponential and power-law in $L$ in the ergodic and MBL phases [Fig.~\ref{fig:xi-dist}(c) and (d)]. The presence of a single lengthscale is also consistent with the scaling of $\braket{(\delta r)^2}$ (defined in Eq.~\ref{eq:rsq-mean}) with $L$. Using Eq.~\ref{eq:xiF-def}, one obtains
\eq{
	\frac{\overline{\braket{(\delta r)^2}}}{L} = \frac{e^{1/\xi_F}}{(1+e^{1/\xi_F})^2}\,.
	\label{eq:rsq-xiF}
}
In the ergodic phase a divergent $\xi_F$  in Eq.~\ref{eq:rsq-xiF} implies $\overline{\braket{(\delta r)^2}}/L=1/4$, which is consistent with the numerical results obtained in Sec.~\ref{sec:fscorr} (see Eq.~\ref{eq:r-erg} and Fig.~\ref{fig:r-rsq}(b)). In the MBL phase the scaling continues to be $\overline{\braket{(\delta r)^2}}\propto L$, as likewise found in Sec.~\ref{sec:fscorr}. 

At the critical point however, it is insufficient to consider $\xi_F$ as $\delta$-function distributed.  Here, while the mode of $P_{\xi_{F}}$ remains of order unity, as seen in Fig.~\ref{fig:r-rsq}(b) the variance $\overline{\braket{(\delta r)^2}}\sim L^a$ with $a>1$, which is in 
qualitative contrast to Eq.~\ref{eq:rsq-xiF}. A super-linear scaling of $\overline{\braket{(\delta r)^2}}$ can however arise from a distribution of $\xi_F$ which is not $\delta$-distributed  in the thermodynamic limit. In such a case, one obtains
\eq{
	\frac{\overline{\braket{(\delta r)^2}}}{L^2}=\int d{\xi_F^{\pd}}&P_{\xi_F}^{\pd}(\xi_{F}^{\pd})\frac{1+L^{-1}e^{1/\xi_F}}{(1+e^{1/\xi_F})^2}  \nonumber\\& -\left(\int d{\xi_F^{\pd}}P_{\xi_F}^{\pd}(\xi_{F}^{\pd})\frac{1}{1+e^{1/\xi_F}}\right)^2\,.
	\label{eq:rsq-Pxi}
}
In general, for a distribution $P_{\xi_F}$  which is not $\delta$-distributed, the right hand side of Eq.~\ref{eq:rsq-Pxi} will be finite, 
whence $\overline{\braket{(\delta r)^2}}\sim L^2$ at the critical point. In contrast, with the distribution $P_{\xi_F}$ a $\delta$-function in Eq.~\ref{eq:rsq-Pxi}, as is the case inside either phase, one recovers Eq.~\ref{eq:rsq-xiF} and hence $\overline{\braket{(\delta r)^2}}\sim L$.

%%%%%%%%%%%%%%%%%%%%%%%%%%%%%%%%%%%%%%%%%%%%%%%%%%%%%%%%%%%%%%%%%%%%%%%%%%%%%%

\subsection{An alternative Fock-space lengthscale \label{sec:xiPi}}
In Sec.~\ref{sec:radialprob} the radial probability distribution $\Pi(r)$~\cite{detomasi2020rare} was discussed. Our physical intuition here is that, on an average, the wavefunction density on Fock-space sites decays exponentially with the distance of the sites from 
the `apex' site $I_0$ (on which the wavefunction has maximal amplitude); i.e.\ that
\eq{
\overline{|A_{K}|^2} = \overline{|A_{I_0}|^2}\exp[{-r_{I_0K}^{\pd}/
\xi_\Pi}^{\pd}]\,,
\label{eq:KW3}
}
which provides us with an alternative lengthscale on the Fock space. $\overline{\Pi(r)}$ (Eq.~\ref{eq:Pir}) can then be recast as 
\eq{
	\overline{\Pi(r)} = N_r^{\pd}\pi(r)\,, 
	\label{eq:pi-def}
}
where $\pi(r)=\overline{|A_{I_0}|^2}\exp[{-r/\xi_\Pi}]$ is the average wavefunction density on a Fock-space site at Hamming distance $r$ 
from $I_0$. Enforcing the (exact) normalisation $\sum_{r=0}^{L}\Pi(r)=1$ gives 
\eq{
\pi(r)=\frac{e^{-r/\xi_{\Pi}^{\pd}}}{(1+e^{-1/\xi_{\Pi}^{\pd}})^L}\,.
\label{eq:xiPi-def}
}
The behaviour of the lengthscale $\xi_\Pi$ is qualitatively similar to that of $\xi_F$, i.e.\ is finite in the MBL phase and divergent in the ergodic phase. In Appendix~\ref{app:mbl0Pi}, we provide exact results for $\Pi(r)$ in the $\mbln$ case, showing in particular that the functional form Eqs.\ \ref{eq:KW3}-\ref{eq:xiPi-def} indeed arises, and obtaining $\xi_{\Pi} \equiv \xi_{\Pi}(W)$.

While $\xi_\Pi$ seems not to be explicitly related to any spatially local autocorrelation or eigenstate expectation value, it can be connected heuristically  to the (real-space) $l$-bit localisation length. Within the $l$-bit picture~\cite{serbyn2013local,huse2014phenomenology}, which is expected to hold at least sufficiently deep in the MBL phase, the MBL eigenstates are understood to be adiabatically connected to the $W\to\infty$ eigenstates via quasilocal unitary transformations: $\ket{E}=U\ket{I_0}$ where $U$ is the quasilocal unitary in question.
The $l$-bit picture also states that the MBL eigenstates are simultaneous eigenstates of dressed versions of the $\sigma^z$-operators, $\tau^z_i = U\sigma^z_i U^\dagger$. It can be shown that the average distance $\braket{d}$, defined in Eq.~\ref{eq:d-dsq} can be expressed as~\cite{detomasi2020rare} 
\eq{
	\frac{\overline{\braket{d}}}{L}= (1+e^{1/\xi_{\Pi}^{\pd}})^{-1}=\frac{1}{2}-\frac{1}{2L}\sum_{i}\overline{\braket{n|\sigma^z_i\tau^z_i|n}}\,.
	\label{eq:d-sigmatau}
}
Heuristically, the $l$-bit operator can be expressed as
\eq{
	\tau^z_i = \frac{1}{\mathcal{N}}\left[\frac{ h_i\sigma^z_i+\Gamma\sigma^x_i}{\sqrt{h_i^2+\Gamma^2}}\mathrm{sgn}(h_i)+\sum_{j\neq i,\mu}J_{j,\mu}e^{-\vert i-j\vert/\zeta}\sigma^\mu_j+\cdots\right]\,
	\label{eq:tauz}
}
where the first term denotes the canting of the spins in the non-interacting limit, and the second term with $\braket{J_{j,\mu}^2}=J^2$ denotes the exponentially decaying support of the $l$-bit operators with a lengthscale $\zeta\equiv\zeta(W,J)$. The normalisation factor 
$\mathcal{N}$, obtained by ensuring that the operator norm of $\tau_i^z$ is conserved, can be estimated as
\eq{
	\mathcal{N}\approx [1+J^2/(1-e^{-2/\zeta})]^{1/2}\,.
}
Using Eq.~\ref{eq:tauz} in Eq.~\ref{eq:d-sigmatau}, an explicit relation between the Fock-space lengthscale $\xi_\Pi$ and the $l$-bit localisation length $\zeta$ can then be obtained as
\eq{
	1+\frac{J^2}{1-e^{-2/\zeta}} = g_W^2\left(\frac{e^{1/\xi_\Pi^{\pd}}+1}{e^{1/\xi_\Pi^{\pd}}-1}\right)^2\,,
	\label{eq:xi-zeta}
}
where $g_W = (\sqrt{W^2+1}-1)/W$ (and which for $J=0$ recovers correctly Eq.~\ref{eq:xipijo} for $\xi_{\Pi}$).

Given that $\xi_\Pi$, like $\xi_{F}$, remains finite in the MBL phase up to the transition, Eq.~\ref{eq:xi-zeta} implies that the $l$-bit localisation length is also finite at the MBL transition, consistent with recent phenomenological and numerical studies~\cite{luitz2017how,thiery2018many-body,morningstar2020manybody}.

%%%%%%%%%%%%%%%%%%%%%%%%%%%%%%%%%%%%%%%%%%%%%%%%%%%%%%%%%%%%%%%%%%

\section{Inhomogeneities on Fock-space \label{sec:inhomo}} 

The correlation function $F(r)$, and the radial probability distribution $\Pi(r)$, probe average properties of the eigenstates on the Fock space. For example, the form of $\pi(r)=\overline{|A_{I_0}|^2}\exp[{-r/\xi_\Pi}]$ implies that the wavefunction in the MBL phase decays exponentially \emph{on an average} with $r$, away from its maximum on the Fock-space site $I_0$. Such measures are however insensitive to the finer stuctures, such as whether or not the exponential decay is homogeneous across all $N_{r}=\binom{L}{r}$ Fock-space sites at given distance $r$ from $I_0$. Besides probing the structure of the wavefunction across the Fock space, the importance of the question lies in the fact that the linear and volumic scaling of the IPRs in the MBL and ergodic phases was motivated from the physical picture that, in the former, eigenstates reside on a sparse network in the Fock space,  whereas in the latter the states are spread homogeneously. In this section, we present results which show this picture to be correct.

To give first a broad qualitative view of the inhomogeneous structure of the eigenstates, we define wavefunction densities $|B_I|^2$, normalised on each row of the Fock-space corresponding to the given distance from $I_0$, as
\eq{
	|B_I|^2 = \frac{|A_I|^2}{\sum\limits_{K: r_{I_0K}^{\pd}=r_{I_0I}^{\pd}}|A_K|^2};\quad |D_I|^2=N_{r_{I_0I}^{\pd}}|B_I|^2\,.
	\label{eq:BIsqDIsq}
}

\begin{figure}
\includegraphics[width=\linewidth]{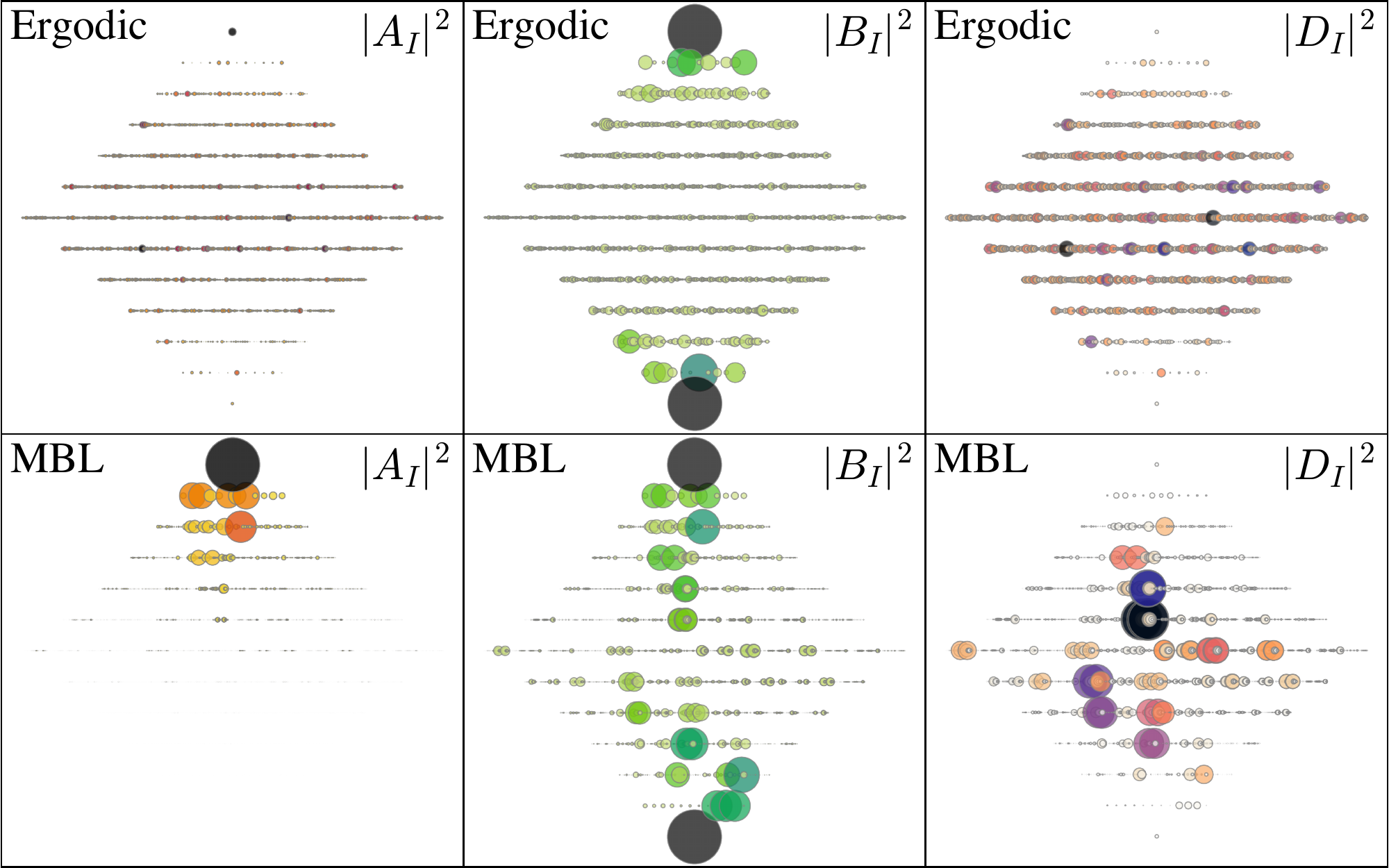}
\caption{Inhomogeneity or its absence in eigenstate densities on the Fock space, in the MBL and ergodic phases respectively. Top and bottom rows correspond respectively to  ergodic ($W=1$) and MBL ($W=6$) phases. In each panel, the corresponding quantity ($|A_I|^2$, $|B_I|^2$, and $|D_I|^2$) is shown on the Fock space (for $L=12$) where the relative size of the blob and its colour denote the value on each Fock-space site (darker colours indicate larger values). The Fock space is arranged such that the site $I_0$ with maximum amplitude is on the top, and all sites on row $r$ lie at distance $r$ from $I_0$.}
\label{fig:inhomo}
\end{figure}

Fig.~\ref{fig:inhomo} shows the numerical distribution of $|A_I|^2$ and $|B_I|^2$ in both the ergodic and MBL phases, with the Fock space arranged such that the site $I_0$ with maximum $|A_{I_{0}}|^{2}$ lies at the top and sites at consecutive rows are at increasing Hamming distance from $I_0$. The relative size denotes the value of $|A_I|^2$ or $|B_I|^2$ on the Fock-space site. In the ergodic phase, homogeneity prevails: the distribution of $|A_I|^2$ is uniform on the Fock space. In the MBL phase by contrast the state is localised near $I_0$,  
with the weights decaying with increasing $r$.

The results for $|B_I|^2$ in the MBL phase render the inhomogeneity clear visually. On each row, there are a few Fock-space sites where the weight is significantly larger than the remaining sites in the row; suggestive of the sparse-network structure of states on the Fock space. In the ergodic phase on the other hand, the $|B_I|^2$ are uniform in the bulk of the Fock space, indicating homogeneity. Note that the weights near both apical sites of the Fock space appear relatively large in either phase; this is simply due to the fact that the number of Fock-space sites on those rows are small (in fact on the apical sites at $r=0$ and $r=L$, Eq.~\ref{eq:BIsqDIsq} forces $|B_I|^2=1$). To distill this effect out, one can consider the wavefunction density on a site $I$ relative to the average density on a site on its row; this is encoded in the quantity $|D_I|^2$ defined in Eq.~\ref{eq:BIsqDIsq}. The results, shown in the third column in Fig.~\ref{fig:inhomo}, are consonant with the picture obtained from $|B_I|^2$. While in the ergodic phase $|D_I|^2$ is homogeneous all over the Fock space, in the MBL phase there are a few sites on each row where the weight is significantly larger than the average weight on that row.

\begin{figure}
\includegraphics[width=\linewidth]{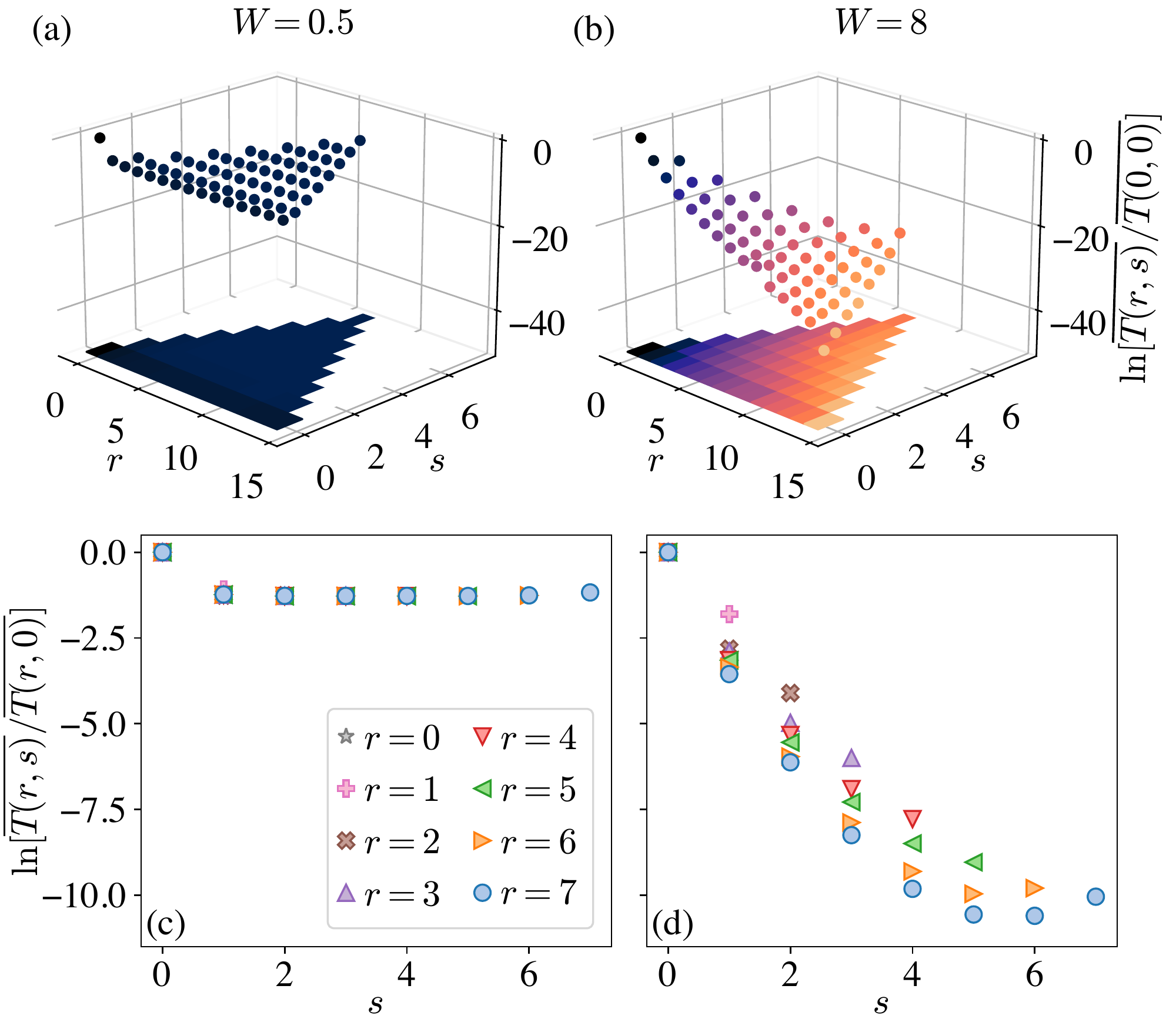}
\caption{The correlation function $T(r,s)$, defined in Eq.~\ref{eq:Trs}, in the ergodic phase [panels (a),(c)] and in the MBL phase [panels (b),(d)]. Panels (a) and (b) show the logarithm of $\overline{T(r,s)}/\overline{T(0,0)}$ as a three-dimensional scatter plot in the space of $r$ and $s$ for $L=14$; the colour-map in the bottom plane shows the same data. Panels (c) and (d) show the logarithm of 
$\overline{T(r,s)}/\overline{T(r,0)}$ as function of $s$ for different $r$. The data in (d) suggests the decay lengthscale in $s$ is 
$r$-independent. }
\label{fig:Trs}
\end{figure}

Having shown qualitatively the inhomogeneity of eigenstates on Fock space in the MBL phase, and lack of it in the ergodic phase, we now discuss quantitative results for them. One measure to quantify the inhomogeneity is the average eigenstate correlation between a pair of Fock-space sites on the same row, $I$ and $K$ such that $r_{I_0I}=r_{I_0K}$. Note that the distance $r_{IK}$ between two such Fock-space sites is necessarily even (there are no links between Fock-space sites on the same row for the model Eq.~\ref{eq:ham}):
$r_{IK}=2s$ where $s=0,1,\cdots,\min[r,L-r]$. The appropriate correlation function is then defined as 
\eq{
	T(r,s)= \frac{1}{N_r N_s^{(r)}}\sum_{\substack{I,K: \\ r_{I_0I}^{\pd}=r=r_{I_0K}^{\pd},\\r_{IK}=2s}}|A_I|^2|A_K|^2\,,
	\label{eq:Trs}
}
where $N_s^{(r)}=\binom{L-r}{s}\binom{r}{s}$ is the number of sites on row-$r$ at Hamming distance $2s$ from any given site on the row. 
The quantity can be interpreted as a higher two-point generalisation of the radial probability distribution. Numerical results for it are shown in Fig.~\ref{fig:Trs}.

In the ergodic phase, the homogeneity of the eigenstates across the entire Fock space leads to the expectation that $T(r,s)$ is independent of $r$ and $s$ and $\sim \nh^{-2}$. This is indeed evident in the results shown in Fig.~\ref{fig:Trs}(a) and (c). In the MBL phase, we expect 
$T(r,s)$ to have an exponentially decaying envelope with $r$, reflecting the exponential decay of the wavefunction densities with $r$. However, a homogeneous decay across all sites at a given $r$ would imply no dependence of $T(r,s)$ on $s$. On the contrary, the results shown in Fig.~\ref{fig:Trs}(b) suggest that $T(r,s)$ decays exponentially with $s$ as well. Indeed the common initial slopes of the results shown in panel (d) suggest that the lengthscale associated with the exponential decay in $s$ is $r$-independent. This implies the form
\eq{
	\overline{T(r,s)} 
	%= \overline{|A_{I_0}|^4}
	~\propto ~ \exp\left(-\frac{r}{\xi_r}\right)\exp\left(-\frac{s}{\xi_s}\right)\,,
	\label{eq:Trsform}
}
where the lengthscales $\xi_r$ and $\xi_s$ are respectively finite and divergent in the MBL and ergodic phases.  Precisely this form for $\overline{T(r,s)}$ can in fact be shown explicitly to arise in  the $\mbln$ case, as considered in Appendix~\ref{app:mbl0trs} where analytical expressions for $\overline{T(r,s)}$ and $\xi_{r/s}$ are obtained. The $W$-dependence of $\xi_{r/s}$  in the $\mbln$ case is qualitatively similar to that of $\xi_F$ and $\xi_\Pi$ (see Fig.~\ref{fig:mbl0xi}).

\begin{figure}
\includegraphics[width=\linewidth]{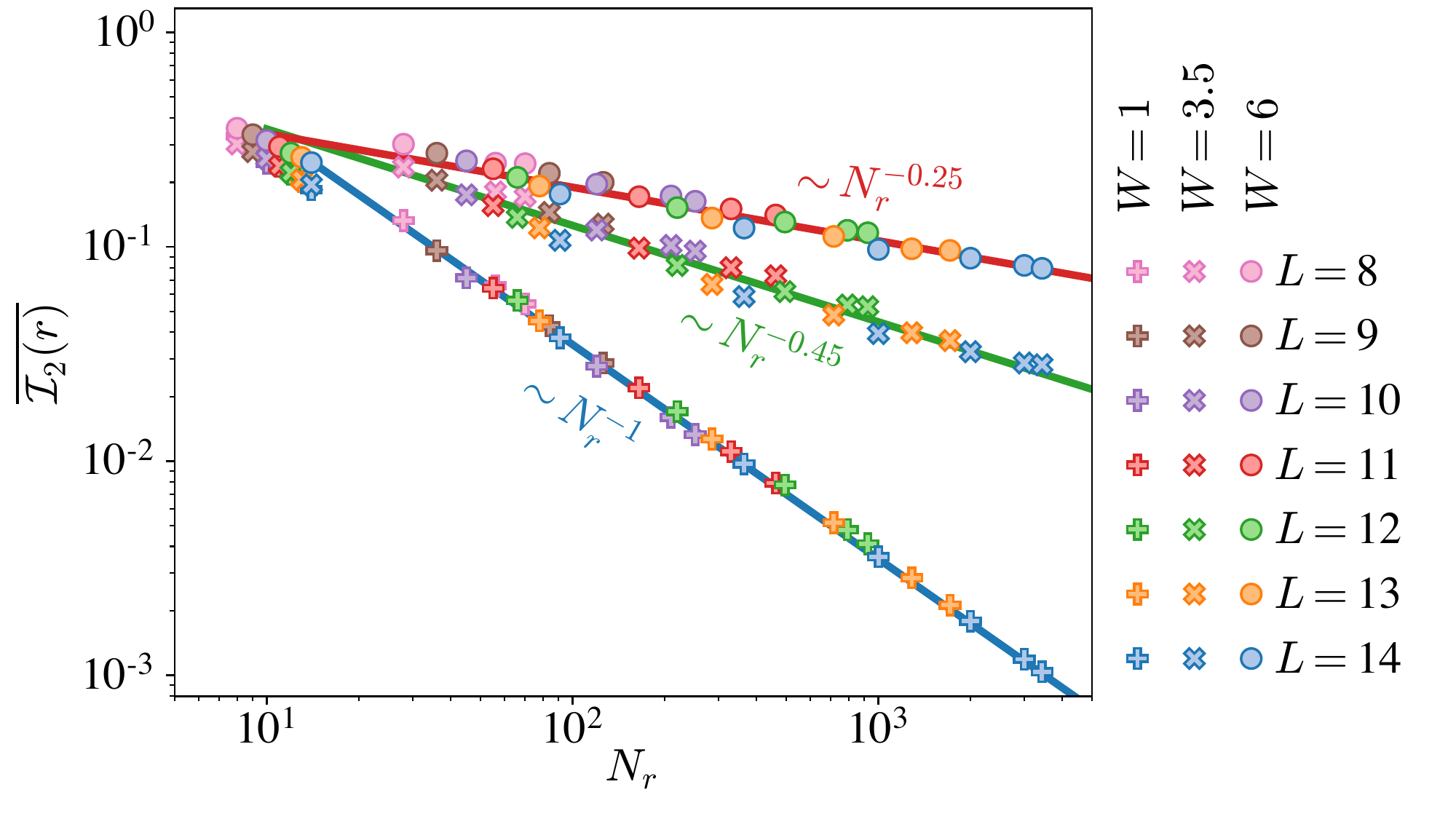}
\caption{Scaling of the distance-resolved IPR, $\mathcal{I}_2$ (Eq.~\ref{eq:I2r}), with the number of Fock-space sites $N_{r}$ at a given distance from the wavefunction maximum. Data are shown for three different disorder strengths (different plot markers), corresponding to the ergodic phase, the critical point, and the MBL phase. The fractal scaling within each row of the Fock-space in the MBL phase and at the critical point, and lack thereof in the ergodic phase, is evident from the scaling. }
\label{fig:I2r}
\end{figure}

A complementary means of quantifying eigenstate inhomogeneity  within a given Fock-space row,  is  via the moments of wavefunction densities 
$|B_I|^2$ (Eq.~\ref{eq:BIsqDIsq}), normalised on each row at a given distance from $I_0$. To this end we define a distance-resolved IPR as
\eq{
	\mathcal{I}_q(r) = \sum_{I: r_{I_0I}=r}|B_I|^{2q}\,
	\label{eq:I2r}
}
(such that by definition $\mathcal{I}_{1}(r)=1 ~\forall r$), and study its scaling with the number of Fock-space sites $N_{r}$ ($=\binom{L}{r}$)
 at distance $r$ from $I_0$: $\mathcal{I}_q(r)\sim N_r^{-\nu_q}$ where the exponent $\nu_q$ is the analogue of the fractal exponent. For a state which is homogeneously distributed across the entire row,  $|B_I|^2\sim N_r^{-1}$ will be independent of $I$ on an average, and hence 
$\nu_q = q-1$.  By contrast, $0<\nu_q<1$ indicates that the wavefunctions show fractal scalings within the rows and are thus inhomogeneously distributed on the rows. We focus on $q=2$, and in Fig.~\ref{fig:I2r} show representative numerical results, in both the ergodic and MBL phases as well as at the critical point. For $W=1$ in the ergodic phase, $\overline{\mathcal{I}_2(r)}\sim N_r^{-1}$ for a range of different $r$ and $L$, showing the wavefunctions to be homogeneously distributed within the rows. By contrast, in the MBL phase and at the critical point,
$\overline{\mathcal{I}_2(r)}\sim N_r^{-\nu_2}$ with $\nu_2<1$, indicating inhomogeneity. We find the exponents $\nu_2$ to be quite close to 
those for the exponent $\tau_2$ characterising the overall mean Fock-space IPR (Sec.~\ref{sec:ipr}). It is reasonable to expect that $\nu_2$ is also discontinuous across the MBL transition, although a large-scale numerical study of the critical behaviour of $\nu_2$ remains for future work.

%%%%%%%%%%%%%%%%%%%%%%%%%%%%%%%%%%%%%%%%%%%%%%%%%%%%%%%%%%%%%%%%%%

\section{Summary and discussion \label{sec:conclusion}}

In summary, we have studied in detail the structure of eigenstates on the Fock space across the MBL transition, via spatial correlations between eigenstate densities on the Fock-space graph. These correlation functions were motivated by their direct connection to eigenstate expectation values, and infinite-time autocorrelation functions, of spatially local observables which diagnose the MBL phase and the MBL transition. 

A central result is that the correlation length associated with the Fock-space correlations changes abruptly across the MBL transition, which in turn implies that the local observables are also discontinuous across the transition. The critical behaviour of the correlation length is obtained from a scaling theory~\cite{mace2019multifractal} based on the Fock-space IPRs, which the correlation functions directly encode. 
The discontinuity in the lengthscale is intimately connected to the discontinuity in the multifractal exponent across the MBL transition~\cite{mace2019multifractal,detomasi2020rare}. Our analysis shows that, throughout the MBL phase and at the MBL transition, the correlation length is finite, which results in the eigenstates exhibiting fractal statistics on the Fock space. Throughout the ergodic phase on the other hand, the multifractal exponent has a value of unity; this is manifest in the correlation length diverging linearly with system size $L$. Importantly, the scaling of the IPRs in the ergodic phase is characterised by a non-ergodicity volume which diverges with an essential singularity at the MBL transition~\cite{mace2019multifractal}. The picture emerging therefore -- that the MBL transition can be interpreted as a terminal end point of a line of fixed points -- suggests that the transition is characterised by a Kosterlitz-Thouless-like scaling, consistent with the predictions from phenomenological treatments in real space~\cite{deroeck2017stability,goremykina2019analytically,dumitrescu2018kosterlitz,thiery2018many-body,morningstar2020manybody}.

We further probed the finer structures of the eigenstates via generalisations of the Fock-space correlations, which showed that MBL eigenstates are strongly inhomogeneous on the Fock space. It is the presence of these inhomogeneities in the MBL phase, and their absence in the ergodic phase, which lies at the heart of the asymmetrical scaling of the Fock-space IPRs in the two phases.

The present work establishes a concrete connection, across the MBL transition, between the behaviour of real-space, local observables and the structure of eigenstates on the Fock space. As such, it naturally provides a pathway towards further study of important questions about the former from the Fock-space perspective, and possibly to unify theories for the MBL transition based on real-space and Fock-space. 
\new{For instance, the Fock-space correlations in this work were motivated by local polarisations averaged over the system. In Ref.~\cite{laflorencie2020chain} on the other hand, extremal values of the local polarisation were considered. Their finite-size scaling was shown to be controlled by a lengthscale which diverges exponentially on approaching the transition from the MBL side. The Fock-space representation of such a quantity and the possibility of a Fock-space correlation length that also diverges exponentially remains an interesting open question.}

While we have here studied  spatial correlations on the Fock space in individual eigenstates, generalising it to  include spatial correlations between eigenstates of different energy -- and hence spatio-temporal correlations on the Fock space which encode the real-time dynamics of local autocorrelations -- is of immediate interest~\cite{creed2021dynamics}.
In this work, within the $l$-bit picture of the MBL phase, we also made a heuristic connection between a Fock-space lengthscale and the $l$-bit localisation length. Several spectral and dynamical properties of the MBL phase are  however determined by rare, spatially local resonances between these $l$-bit integrals of motion~\cite{gopalakrishnan2015lowfrequency,villalonga2020eigenstates,morningstar2021avalanches,garratt2021resonances}. Understanding how such resonances are manifest in the inhomogeneous structure of the MBL eigenstates on the Fock space is a question of
evident interest.

Finally, we have considered a system where the disordered real-space fields and couplings are uncorrelated random variables. In such systems, the absence of self-averaging and broad distributions in certain quantities near the transition, as we observed, may be attributed to Griffiths effects~\cite{agarwal2015anomalous,gopalakrishnan2016griffiths}. Such effects are however absent in MBL systems with quasiperiodic Hamiltonians; indeed distributions of quantities that are broad near the MBL transition in disordered systems typically remain narrow in quasiperiodic systems~\cite{sierant2019level}. The nature of the MBL transition may thus be different in quasiperiodic 
systems~\cite{khemani2017two}. It would therefore be worthwhile to develop a scaling theory of the MBL transition in quasiperiodic systems, and 
to understand the behaviour of local observables along the lines discussed in this work.

%%%%%%%%%%%%%%%%%%%%%%%%%%%%%%%%%%%%%%%%%%%%%%%%%%%%%%%%%%%%%%%%%%
\begin{acknowledgments}
We are grateful for helpful discussions with 
S.~Banerjee, J.~Bon\v{c}a, J.~T.~Chalker,  I.~Creed, G.~De Tomasi, A.~Duthie, S.~J.~Garratt, I.~Khaymovich, A.~Lazarides,  M. Mierzejewski, S.~Mukerjee and L. Vidmar.  This work was in part supported by EPSRC Grant No. EP/S020527/1.
\end{acknowledgments}

%%%%%%%%%%%%%%%%%%%%%%%%%%%%%%%%%%%%%%%%%%%%%%%%%%%%%%%%%%%%%%%%%%%%

\appendix
\section{Analytical results deep in the MBL phase \label{sec:non-int-corr}}

In this appendix we provide some analytical results valid at strong disorder deep in the MBL phase. In this regime the MBL phase is adiabatically connected to the non-interacting limit of $J_i=0$, such that the model describes a collection of non-interacting spins and 
is thus `trivially' MBL. Nevertheless, the tensor-product structure of the Fock space still renders non-trivial many properties of the eigenstates on the Fock space. This limit faithfully captures certain aspects of the genuine MBL phase in the presence of interactions,
in particular the scaling of several quantities on the Fock space with system size.

The Hamiltonian in this case, referred to as $\mbln$ in the text, is
\eq{
	H_{\mbln} = \sum_i [\sigma^x_i + h_i^\pd\sigma^z_i]\,.
}
Its eigenstates are product states in the $\tau^z$-basis, with the $\tau^z$-operators given by
\eq{
	\tau^z_i = \frac{h_i\sigma^z_i+\sigma^x_i}{\sqrt{h_i^2+1}}\mathrm{sgn}(h_i)\,.
}
An eigenstate $\ket{E}$ can be labelled by its configuration of $\tau$-spins, $\ket{E}=\ket{\tau_1,\tau_2,\cdots,\tau_L}$, where 
$\tau_i=\pm 1$. With this notation, the overlap of the eigenstate with a Fock-space site $I$ (in the $\sigma^z$-basis) can be expressed as
\eq{
	\vert A_{I}\vert^2 = \prod_{i=1}^L p_{i,I}\,
	\label{eq:mbl0AI}
}
where 
\eq{
p_{i,I} =p_{i}^+\delta_{S_i^I,+\tau_i}^{\pd}+p_{i}^-\delta_{S_i^I,-\tau_i}^{\pd}
	\label{eq:mbl0Apdef}
}
with
\eq{
	p_{i}^\pm= \frac{1}{2}\Big(1\pm\frac{\vert h_i\vert}{\sqrt{h_i^2+1}}\Big).
	\label{eq:wombat}
}
It will  prove useful to define the following notation for the ($i$-independent) disorder averages:
\eq{
	&p_{\pm} \equiv \overline{p_i^\pm} = [W\pm(\sqrt{W^2+1}-1)]/2W\,\label{eq:ppm}\\
	&q_\pm \equiv \overline{(p_i^\pm)^2}= \frac{1}{2}\left(1-\frac{\tan^{-1}W}{2W} \pm \frac{\sqrt{W^2+1}-1}{W}\right)\,\label{eq:qpm}\\
	&q\equiv\overline{p_i^+p_i^-}=(\tan^{-1}W)/4W\,.\label{eq:q}
}

%%%%%%%%%%%%%%%%%%%%%%%%%%%%%%%%%%%%%%
\subsection{Spatial correlations in Fock space \label{app:mbl0Fr}}

The  correlation function $\Fr$ in the $\mbln$ case can be computed as follows. For two Fock-space sites $I$ and $K$ such that $r_{IK}=r$, there are by definition $r$ real-space sites with $S_i^I=-S_i^K$, while the remaining  $L-r$ real-space sites have $S_i^I=S_i^K$.
On the former category of real-space sites, we have $p_{i,I}=p_i^+$ and $p_{i,K}=p_i^-$, or $p_{i,I}=p_i^-$ and $p_{i,K}=p_i^+$. 
For the latter category, we have $p_{i,I}=p_i^+=p_{i,K}$, or $p_{i,I}=p_i^-=p_{i,K}$. Hence, on the average, 
\eq{
	\overline{|A_I|^2|A_K|^2} = (q^r) [(q_++q_-)/2]^{L-r}\,,
}
such that 
\eq{
	\overline{F_{\mbln}(r)} = \nh N_r (q^r) [(q_++q_-)/2]^{L-r}\,.
	\label{eq:mbl0another}
}
With $\nh =2^{L}$, and noting that $q_{+}+q_{-} =1-2q$, Eq.~\ref{eq:mbl0another} rearranges to:
\eq{
	\overline{F_{\mbln}(r)} = N_r (1-2q)^{L}\exp\left[-r\ln\left(\frac{1-2q}{2q}\right)\right]\,
}
This is precisely Eq.~\ref{eq:F-MBL*}, recognising (Eq.~\ref{eq:q}) that $2 q\equiv p$ as defined therein. The lengthscale $\xi_F$ can
thus be identified as $\xi_F^{-1}=\ln[(1-p)/p]$. Fig.~\ref{fig:mbl0xi} shows the resultant behaviour of $\xi_F$ with disorder strength $W$.

%%%%%%%%%%%%%%%%%%%%%%%%%%%%%%%%%%%%%%%%%

\subsection{IPR exponent and fluctuations \label{app:mbl0ipr}}
For the $\mbln$ case, the IPR $\mathcal{L}_2 = \sum_{I}|A_I|^4$ of the eigenstate can be expressed as
\eq{
	\mathcal{L}_2 =
	\prod_{i=1}^{L} \sum_{s_{i}^{I}=\pm 1}p_{i,I}^{2} =\prod_{i=1}^L[(p_i^+)^2+(p_i^-)^2]
}
which for the average leads to $\ipr = (1-p)^L$ with $p=2q =(\tan^{-1}{W})/2W$, and hence $\tau_2 = -\ln(1-p)/\ln 2$. Noting that
$\tau_2 \sim -\ln\ltwo/\ln\nh$ for a given realisation, the variance of $\tau_2$ can be obtained as 
\eq{
	\braket{(\delta \tau_2)^2}=\frac{1}{L}\frac{u}{(\ln 2)^2},
}
where $u = \braket{\ln^2[1+h^2/(h^2+1)]}-\braket{\ln[1+h^2/(h^2+1)]}^2$. This $1/L$ decay of $\braket{(\delta \tau_2)^2}$ persists in the interacting MBL phase as well at strong disorder, see Fig.~\ref{fig:tau-dist}(d). 

%%%%%%%%%%%%%%%%%%%%%%%%%%%%%%%%%%%%%%%%%

\subsection{Radial probability distribution and associated lengthscale \label{app:mbl0Pi}}
\begin{figure}
\includegraphics[width=0.75\linewidth]{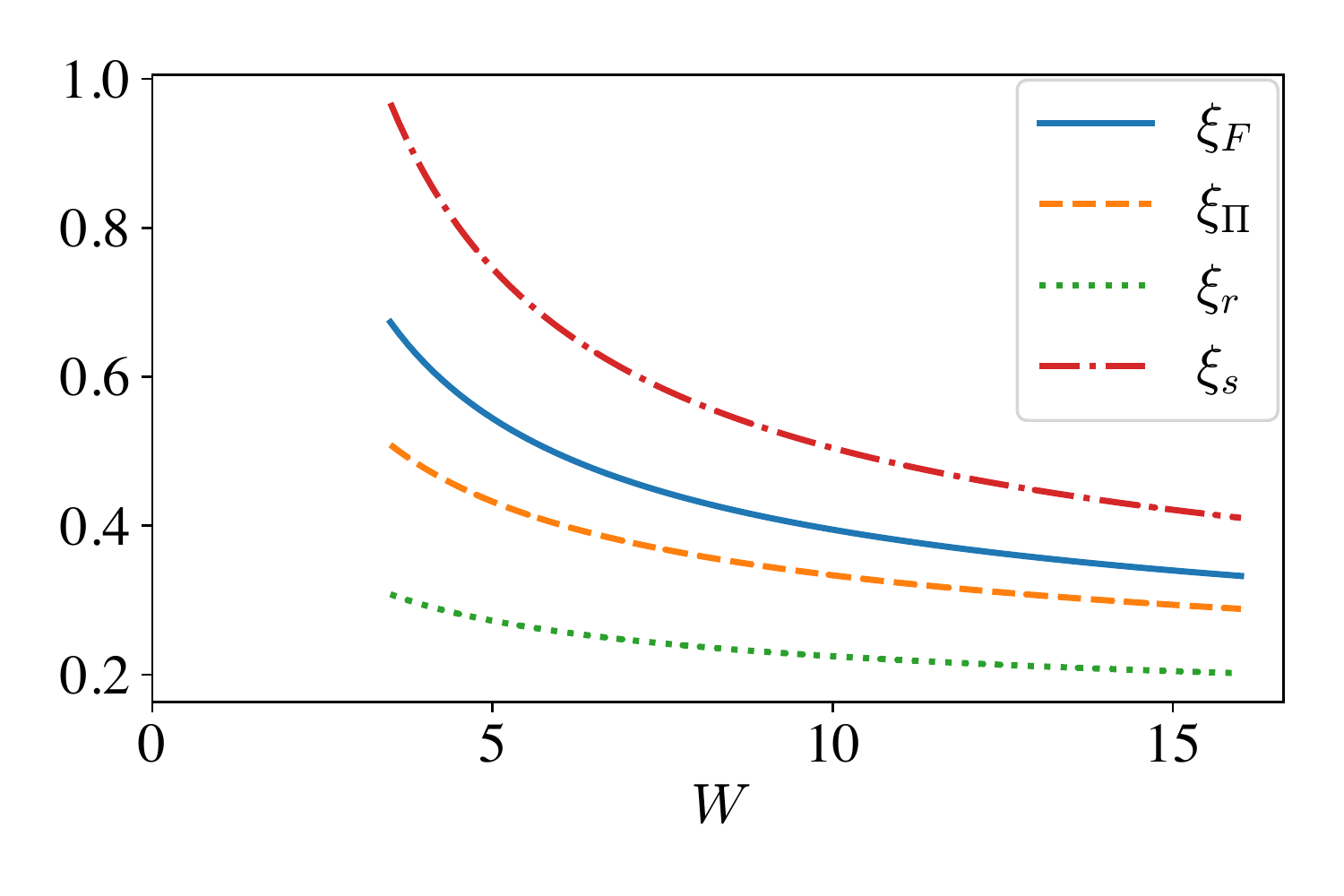}
\caption{$W$-dependence of the four Fock-space lengthscales in the $\mbln$ case. All have qualitatively similar behaviour.
}
\label{fig:mbl0xi}
\end{figure}

The radial probability distribution in the $\mbln$ case can be obtained as follows. Given an eigenstate $\ket{E}=\ket{\{\tau_i\}}$, 
and noting (Eq.~\ref{eq:wombat}) that $p_{i}^{+}>p_{i}^{-}$, the Fock-space site $\ket{I_0}$ for which the wavefunction amplitude is maximal 
is uniquely identified as that for which $\tau_i=S_{i,I_0}$ for all real-space sites $i$.  One thus has
\eq{
	\overline{\vert A_{I_0}\vert^2} = \prod_{i=1}^L \overline{p_i^+} = (p_+)^L\,,
}
and for any Fock-space site $K$ such that $r_{I_0K}=r$,
\eq{
	\overline{\vert A_{K}\vert^2} = (p_-)^r(p_+)^{L-r}\,.
	\label{eq:ank-nonint}
}
From Eq.~\ref{eq:ank-nonint}, $\overline{\Pi(r)}$ (Eq.~\ref{eq:Pir}) follows directly as
\eq{
	\overline{\Pi(r)} = N_r(p_-)^r(p_+)^{L-r}=N_r(p_+)^Le^{-r\ln\left(\frac{p_+}{p_-}\right)}\,.
}
This is indeed precisely of form Eqs.~\ref{eq:pi-def},\ref{eq:xiPi-def}, with $\xi_\Pi$ given explicitly by
\eq{
	\xi_\Pi^{-1} = \ln\left(\frac{p_+}{p_-}\right)=\ln\left[\frac{W+(\sqrt{W^2+1}-1)}{W-(\sqrt{W^2+1}-1)}\right]\,
	\label{eq:xipijo}
}
(using Eq.~\ref{eq:ppm} for the second equality).
The lengthscales $\xi_\Pi$ and $\xi_F$ are compared in Fig.~\ref{fig:mbl0xi}, and their behaviours with $W$  are seen to be qualitatively similar.

%%%%%%%%%%%%%%%%%%%%%%%%%%%%%%%%%%%%%%%%%

\subsection{Correlation function $T(r,s)$ \label{app:mbl0trs}}
The correlation function $T(r,s)$ defined in Eq.~\ref{eq:Trs} can be obtained in the $\mbln$ case using Eqs.~\ref{eq:mbl0AI},\ref{eq:mbl0Apdef}
as follows. The Hamming distance between two Fock-space sites $I$ and $K$ such that $r_{I_0I}=r=r_{I_0K}$ is necessarily even, $r_{IK}=2s$, due to topology of the graph associated with the Hamiltonian Eq.~\ref{eq:ham}. $r$ real-space sites in $I$ have $s_{i}^{I}=-\tau_{i}$, likewise
$r$ sites in $K$ have $s_{i}^{K}=-\tau_{i}$. Of these two overlapping sets of $r$ sites, $r-s$ of them  must have $s_{i}^{I}=s_{i}^{K}$ $(=-\tau_{i})$ since the Hamming distance $r_{IK}=2s$, and hence have $p_{i,I}p_{i,K}=(p_i^-)^2$. It follows that $2s$ sites then  have $s_{i}^{I}=-s_{i}^{K}$ $(=\pm\tau_{i})$ and hence $p_{i,I}p_{i,K}=p_i^-p_i^+$, while the remaining $L-r-s$ real-space sites have $s_{i}^{I}=s_{i}^{K}=+\tau_{i}$, and thus have $p_{i,I}p_{i,K}=(p_i^+)^2$.  Noting Eqs.\ \ref{eq:qpm},\ref{eq:q} we therefore have, for the average,
\eq{
	\overline{T(r,s)}=\overline{|A_I|^2|A_K|^2} = (q^{2s})(q_+^{L-r-s})(q_-^{r-s})\,.
}
This can be recast as 
\eq{
	\overline{T(r,s)}=(q_+^L)\exp\left[-r\ln\left(\frac{q_+}{q_-}\right)\right]\exp\left[-s\ln\left(\frac{q_+q_-}{q^2}\right)\right]\,
}
(where the $r=0=s$ limit yields $(q_{+})^L=\overline{|A_{I_{0}}|^{4}}$). This is precisely of the form Eq.~\ref{eq:Trsform}, and
leads to the identification of the lengthscales $\xi_r$ and $\xi_s$ as 
\eq{
	\xi_r^{-1}=\ln(q_+/q_-);\quad\xi_s^{-1}=\ln(q_+q_-/q^2)\,,
}
with the relevant expressions given in Eqs.~\ref{eq:qpm} and \ref{eq:q}.  As seen in Fig.~\ref{fig:mbl0xi}, the qualitative behaviour of 
$\xi_r$ and $\xi_s$ is the same as that of $\xi_F$ and $\xi_\Pi$.

%%%%%%%%%%%%%%%%%%%%%%%%%%%%%%%%%%%%%%%%%

\bibliography{refs}
\end{document}